\documentstyle[aps,epsfig]{revtex}

\newcommand{\mb}[1]{ {\mbox{\boldmath{$#1$}}}  }
\begin{document}


\setcounter{footnote}{1}
\title{BCS to Bose Crossover in Anisotropic Superconductors.} 
\author{J.P.Wallington and James F.Annett.}
\address{University of Bristol, H.H. Wills Physics Laboratory,\\
         Royal Fort, Tyndall Ave,\\
         Bristol BS8 1TL, United Kingdom.}
\date{\today}
\maketitle

\begin{abstract}

In this work we use functional integral techniques to examine the nearest
neighbour attractive Hubbard model on a quasi-2D lattice.  It is a simple
phenomenological model for the high-T$_c$ cuprates that allows both
extended (non-local) $s$- and $d$-wave singlet superconductivity as well as
mixed symmetry states.  The Hartree-Gor'kov mean field theory of the model
has a finite temperature phase diagram which shows a transition from pure
$s$-wave to pure $d$-wave superconductivity, via a mixed symmetry $s+id$
state, as a function of doping.  Including Gaussian fluctuations we examine
the crossover from weak-coupling BCS superconductivity to the
strong-coupling Bose-Einstein condensation of composite $s$- or $d$-wave
bosons and comment on the origin and symmetry of the pseudogap.
\end{abstract}



\section{Introduction}


The physics of the high-T$_c$ cuprates is unusual and fascinating for
a number of reasons.  In this work we explore two of them: the
symmetry of the superconducting pairing state and the existence of the
pseudogap state between the insulator and the superconductor.  

In recent years the experimental evidence has become overwhelming that the
cuprate high-T$_c$ superconductors possess an anisotropic $d$-wave pairing
state\cite{AnnettGL96}.  Angle resolved photoemission (ARPES) experiments
\cite{Loeser96,Ding96} and measurements of the temperature dependence of
the penetration depth \cite{Hardy93,Panagopolous96,Lee96} both strongly
support this scenario, but perhaps the most convincing evidence comes from
the elegant measurements of Wollman {\it et al.}\cite{Wollman93} and of
Tsuei {\it et al.}\cite{Tsuei94} since they do not depend on the
microscopic physics of the energy gap, but instead depend only on the order
parameter phase.  

Whilst there is a consensus on the $d$-wave nature of the dominant part of
the order parameter, there is continuing controversy over the possible
existence of a sub-dominant $s$-wave contribution \cite{Kelley96,Klemm96}.
The existence, or otherwise, of a subdominant $s$-wave component has
profound consequences for the nature of the microscopic pairing mechanism.
For example, antiferromagnetic spin fluctuations lead to attraction in the
$d_{x^2-y^2}$ pairing channel, but are pair breaking in the $s$-wave
channel \cite{Monthoux91}. Similarly the Hubbard model with a positive
(repulsive) on-site interaction U may have a $d_{x^2-y^2}$ paired ground
state, but would presumably not support $s$-wave Cooper pairs.  In the case
of YBa$_2$Cu$_3$O$_7$ the orthorhombic crystal symmetry makes some non-zero
$s$-wave component inevitable, but a large $s$-wave component would be
difficult to reconcile with either of these pairing mechanisms. On the
other hand, pairing mechanisms based on electron-phonon interactions,
polarons, or other non-magnetic excitations (e.g. excitons, acoustic
plasmons) could be compatible with either $s$-wave or $d$-wave pairing
states \cite{Varelogiannis98}.  In these models, whether $s$-wave, $d$-wave
or mixed pairs are more strongly favoured would depend on details of the
model parameters, and could even vary from compound to compound.  Indeed
there is some evidence that the n-type cuprate superconductors are
$s$-wave\cite{Anlage94} (or at least they have no zeros in the gap
$|\Delta({\bf k})|$ on the Fermi surface).  This would imply that either
the pairing mechanism is different for the n- and p-type materials, or that
the mechanism allows both $s$-wave or $d$-wave ground states depending on
the band filling.  

The second remarkable feature of the cuprates is the ``pseudogap'' occuring
in underdoped materials.  Transport experiments \cite{Batlogg94}, ARPES
\cite{Loeser96,Ding96}, NMR \cite{Warren89,Takigawa91}, and optical
conductivity \cite{Homes93,Puchkov96} all detect a loss of spectral weight
at low energies in the one-particle excitation spectrum below a certain
temperature, ${\rm T}^*>{\rm T}_c$.  Furthermore, the ${\bf k}$-specific
nature of the ARPES technique has shown that the pseudogap has nodes at the
same places as the superconducting gap.  These data and the fact that the
coherence length is known to be very small in the cuprates ($\xi(0) \simeq
10-20$ {\AA} in the CuO$_2$ planes) points to a strong-coupling scenario in
which superconducting fluctuations play a major role.

Although there is as yet no concensus on the exact nature of the
fluctuations, it is generally believed that the opening of the pseudogap at
T$^*$ is due to the formation of an incoherent pairing state above T$_c$.  
 
Two competing scenarios have emerged to explain this phenomenon.  In the
first, exemplified by the work of Emery and Kivelson
\cite{EmeryKivelson95}, the relevant fluctuations are in the phase.  The
stiffness to the phase fluctuations is $n_s(0)/m^*$ (where $n_s(0)$ is the
zero temperature superfluid density and $m^*$ is the effective
quasiparticle mass), and where this is small, such as in the underdoped
cuprates, the true T$_c$, where phase coherence is established, can be well
below the mean field transition temperature which is identified with T$^*$.

\begin{figure}
\centerline{\epsfig{file=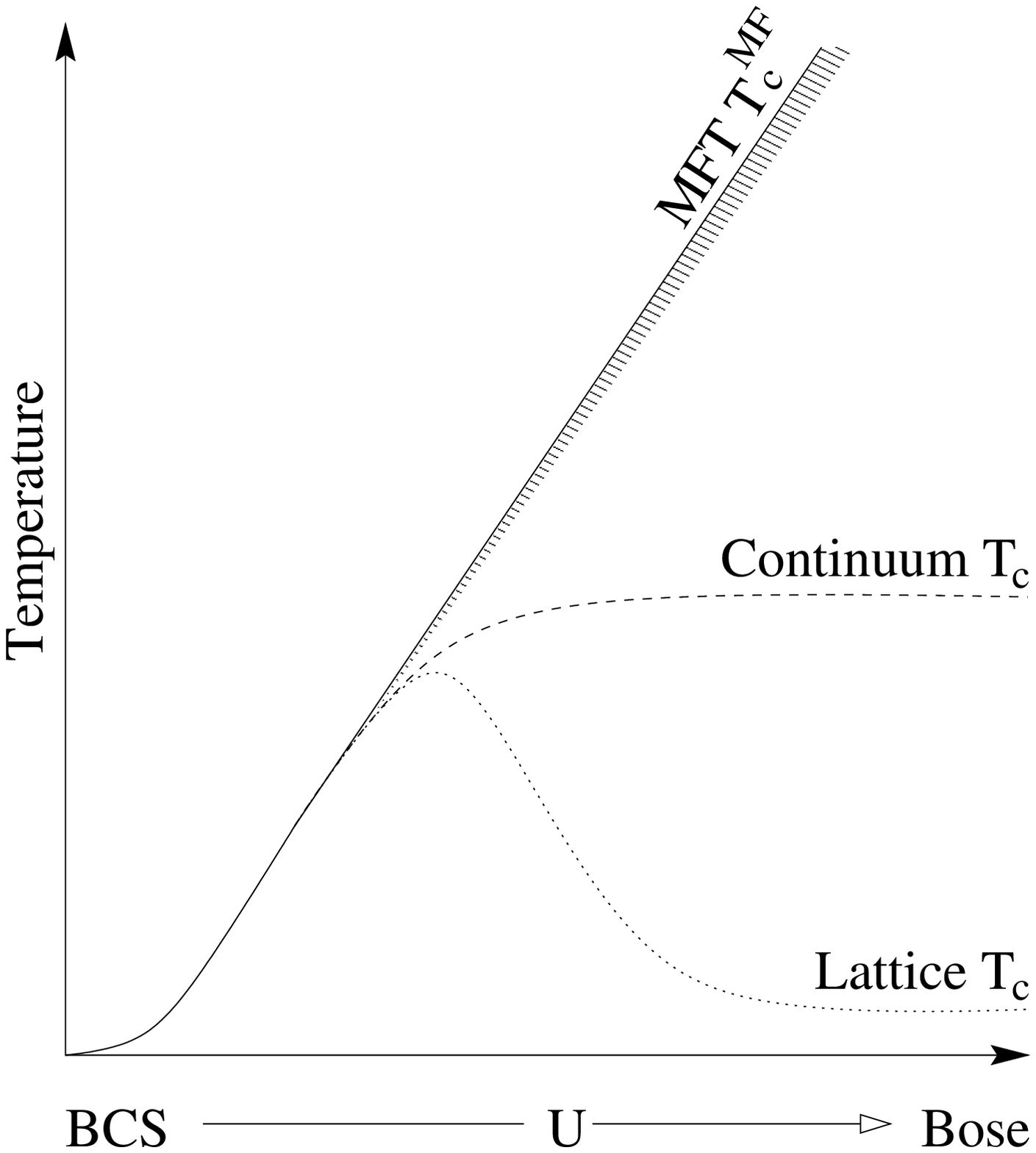,width=7.0cm,angle=0}}
\caption{Crossover from BCS to Bose pairing for continuum and lattice
models.  The full line shows the BCS mean field T$_c$ evaluated for all
strengths of the coupling, $U$.  The broken lines show the effects of
fluctuations on T$_c$.  }
\label{bcs_bose}
\end{figure}

The second scenario is best illustrated by the works of Nozi{\` e}res and
Schmitt-Rink \cite{NSM85} (which predates the discovery of the high-T$_c$
compounds), Dreschler and Zwerger \cite{DreschlerZwerger92} and S{\'a} de
Melo, Randeria and Engelbrecht \cite{SadeMeloRanderiaEngelbrecht93} (see
Ref. \cite{Randeria95} for a review).  By considering both magnitude
and phase fluctuations in $s$-wave superconductors in the vicinity of the
critical temperature, these authors examined the crossover from BCS-like
superconductivity of weakly bound pairs to the Bose condensation of tightly
bound pairs as $U$, the strength of the interaction, was varied.  The
results are summarised in Fig. \ref{bcs_bose}.  For weak pairing the
fluctuating and mean field theories are essentially the same.  However as
$U$ is increased large differences become apparent: whilst the mean field
critical temperature, T$_c^{MF}$, keeps getting larger, the effects of
fluctuations cause the true T$_c$ to flatten off, for the continuum model,
or even to decrease in the case of the lattice model.  The difference
between T$_c^{MF}$ and T$_c$ gives an estimate for the regime where
fluctuations dominate and may correspond to the pseudogap regime in which
bosonic pairs exist but are not condensed into a superfluid.

In this paper we will address both the pairing symmetry and the fluctuation
issues in a systematic way.  We broadly follow the functional integral
approach taken by Dreschler and Zwerger \cite{DreschlerZwerger92} and
S{\'a} de Melo, Randeria and Engelbrecht
\cite{SadeMeloRanderiaEngelbrecht93} , but explicitly including $d$-wave
pairing.  In section \ref{sect_LGW} we introduce the attractive nearest
neighbour Hubbard model as a simple phenomenological model of the cuprates
and derive from it a Landau-Ginzburg-Wilson (LGW) functional describing
$s$- and $d$-wave superconducting fields.  In section \ref{sect_MFT} we
examine the BSC-like mean field theory of the model by finding the saddle
point solutions of the LGW functional which show $s$- and $d$-wave and
mixed symmetry phases.  Finally, in section \ref{sect_GFA}, we include
Gaussian fluctuations about the saddle point in order to study the strong
coupling bosonic limit.  This limit is especially interesting in this model
because of the competition between preformed pairs of different symmetries.
We also comment on the symmetry and origin of the pseudogap.


\section{Landau-Ginzburg-Wilson theory}
\label{sect_LGW}


To investigate the symmetry of the order parameter and the nature of the
bosonic limit for anisotropic pairs, we consider a simple phenomenological
model of the cuprates, the attractive nearest neighbour Hubbard model on a
2D square lattice.  It is the simplest model that allows $s$-wave, $d$-wave
or mixed symmetry pairing states to occur.  Its Hamiltonian is,
\begin{equation}
\hat{H} = \hat{H}_0 + \hat{H}_I,
\end{equation}
where $\hat{H}_0$ contains the chemical potential and a nearest neighbour
hopping term.  The interaction Hamiltonian describes an electron-electron
attraction on nearest neighbour sites:
\begin{equation}
\hat{H}_I =
- V \sum_{\langle ij \rangle } \hat{n}_{i} \hat{n}_{j},
\end{equation}
where $\sum_{\langle ij \rangle}$ denotes summation over all of the bonds
between the nearest neighbour sites.  For the sake of simplicity, we do not
include an on-site interaction term.

In order to investigate the thermodynamics of this model it is sufficient
to consider the partition function.  In the coherent state functional
integral representation it is,
\begin{eqnarray}
{\cal Z}
&=& \int {\cal D}[c^{\dagger},c^{}]
    \;e^{{\cal S}},\\
{\cal S}
&=& \int_{0}^{\beta}d\tau\;
    \Bigg[
        \sum_{i\sigma}  
                c^{\dagger}_{i\sigma}(\tau) 
                \left(
                    -\partial_\tau
                \right) 
                c^{}_{i\sigma}(\tau)
        - H(c^{\dagger},c)
    \Bigg], \nonumber
\end{eqnarray} 
where $c$ and $c^{\dagger}$ are the Grassmannian counterparts to the Fermi
operators $\hat{c}$ and $\hat{c}^{\dagger}$.

Unfortunately, the interaction part of the Hamiltonian is quartic in the
Grassmann (fermion) fields so the functional integral cannot be evaluated
exactly.  However, by using the Hubbard-Stratonovi{\v c} transformation
\cite{Hubbard_HST,Stratonovic_HST} the quartic problem of interacting
electrons can be converted to the more tractable quadratic problem of
non-interacting electrons coupled to a random (Gaussian weighted) Bose
field.  In order to make this transformation, the electronic Hamiltonian
must first be decomposed into a product of fermionic bilinears.  This can
be done in a number of different ways, but the purpose of this paper is to
study superconductivity so we choose what might be called the Gor'kov
decomposition in terms of pairing operators:  
\begin{eqnarray}
\hat{H}_I &=& -V\sum_{\langle ij \rangle}
               {\rm tr}\left(\hat{F}^{\dagger}_{ij}\hat{F}_{ij}\right),
\end{eqnarray}
with the pairing matrix written as,
\begin{eqnarray}
\hat{F}_{ij} &=& \frac{i\sigma_y}{\sqrt{2}}
           \left(  
           \hat{B}_{ij} - \mb{\sigma}\cdot\hat{{\bf B}}_{ij}
           \right),
\end{eqnarray}
where $\mb{\sigma}$ is the vector of Pauli matrices, and
\begin{eqnarray}
\hat{B}_{ij}       &=& \frac{1}{\sqrt{2}}
                        \left(
                        \hat{c}_{j\uparrow} \hat{c}_{i\downarrow}
                        -\hat{c}_{j\downarrow} \hat{c}_{i\uparrow}
                        \right),\\
\hat{{\bf B}}_{ij} &=& 
    \left\{
    \begin{array}{l}
        \frac{1}{\sqrt{2}}
            \left(
                \hat{c}_{j\downarrow} \hat{c}_{i\downarrow}
                -\hat{c}_{j\uparrow} \hat{c}_{i\uparrow}
            \right),\\
        \frac{i}{\sqrt{2}}
            \left(
                \hat{c}_{j\downarrow} \hat{c}_{i\downarrow}
                +\hat{c}_{j\uparrow} \hat{c}_{i\uparrow}
             \right),\\
        \frac{1}{\sqrt{2}}
            \left(
                \hat{c}_{j\uparrow} \hat{c}_{i\downarrow}
                +\hat{c}_{j\downarrow} \hat{c}_{i\uparrow}
            \right).
    \end{array}
    \right.
\end{eqnarray}
The scalars, $\hat{B}_{ij}$, are even under both time reversal and parity whilst
the vectors, $\hat{{\bf B}}_{ij}$, are odd under both.  In the manner of Volovik
and Gor'kov \cite{VolovikGorkov85}, we interpret $\hat{B}_{ij}$ and $\hat{{\bf
B}}_{ij}$ as the annihilation operators for singlet and triplet Cooper pairs
on the bonds $\langle ij \rangle$.  In terms of these operators the
interaction Hamiltonian becomes:
\begin{equation}
\hat{H}_I = -V\sum_{<ij>}
                \left(
                    \hat{B}_{ij}^{\dagger} \hat{B}_{ij}
                    +\hat{{\bf B}}_{ij}^{\dagger}\cdot\hat{{\bf B}}_{ij}
                \right), \label{hamtwo} \label{hami}
\end{equation}
where we have used tr$(\sigma_i)=0$ and tr$(\sigma_i\sigma_j)=2\delta_{ij}$.

Although it would be interesting to study triplet superconductivity,
especially in light of the exciting recent discovery \cite{Luke98} of
$p$-wave pairing in strontium ruthenate, Sr$_2$RuO$_4$, the focus of this
work is on anisotropic {\em singlet} pairing as seen in the high-T$_c$
cuprates.  Therefore, for the sake of simplicity, we restrict our attention
to the singlet fields.

Having decomposed the interaction Hamiltonian into a product of fermionic
bilinears we make the Hubbard-Stratonovi{\v c} transformation.  Tensor
valued Bose fields, $\Delta_{ij}$, are introduced conjugate to the fermion
bilinears, $\hat{B}_{ij}$. (Had we kept the triplet operators we would have needed
to introduce an additional vector of triplet pairing fields, ${\mb
\Delta}_{ij}$.)  In terms of these the partition function becomes,
\begin{eqnarray}
{\cal Z} 
&=&  \int {\cal D}[\Delta^*,\Delta]
        \;e^{S}, \\
S 
&=&  - V \int_{0}^{\beta} {\rm d}\tau \sum_{\langle ij \rangle }
       |\Delta_{ij}(\tau)|^2 
    + \ln{\int {\cal D}[c^{\dagger},c] \; e^{{\cal S}_f}},\nonumber
\end{eqnarray}
with the linearised fermions contained in,
\begin{eqnarray}
{\cal S}_f
&=& \int_{0}^{\beta}{\rm d}\tau\;
    \Bigg[
        \sum_i  c^{\dagger}_{i}(\tau) 
                \left(
                    -\partial_\tau 
                \right) 
                c^{}_{i}(\tau)
        - H_0(c^{\dagger},c) \nonumber \\
&& \hspace{3.5cm}
        - V\sum_{\langle ij \rangle }
            \left(
                \Delta_{ij}^*(\tau) B_{ij}(\tau) 
              + \Delta_{ij}(\tau) B^{\dagger}_{ij}(\tau)
            \right)
    \Bigg].\nonumber
\end{eqnarray}
The Hubbard-Stratonovi{\v c} fields, $\Delta_{ij}$, are defined on each bond $\langle ij
\rangle $ of the lattice.  However, it is generally more convenient to work
with site-centred combinations possessing a definite symmetry.  On a square
lattice, every site, $i$, has two fields associated with it, one in the
$x$-direction and one in the $y$-direction. (We could associate four fields
with each point but that introduces problems with double counting.)  Rather
than talk about $x$- and $y$-bond fields, it is natural to form linear
combinations of them:
\begin{eqnarray}
\Delta_s  &=&
    \frac{1}{2} (\Delta_x + \Delta_y),\\
\Delta_d  &=& 
    \frac{1}{2} (\Delta_x - \Delta_y),
\end{eqnarray}
where $\Delta_a({\bf r}_i)=\Delta_{ii+a}$ for $a=x,y$ according to the
direction of the bond $\langle ij \rangle$.  We call these the $s$- and
$d$-wave fields because, in the spatially homogeneous limit, they become
the order parameters for (extended) $s$- and $d$-wave superconductivity.
Unlike the $x$ and $y$ fields, the $s$ and $d$ order parameters transform
as irreducible representations of the tetragonal crystallographic point
group, $D_{4h}$.  In the notation of Ref. \cite{Annett90} they belong
to the $A_{1g}$ and $B_{1g}$ representations of $D_{4h}$.  The $d$-wave
representation, $B_{1g}$, corresponds to an order parameter with $x^2-y^2$
symmetry as seen in the cuprate materials.  

Proceeding with the calculation, the action is diagonalised by the
introduction of Fourier transformed fields:
\begin{eqnarray}
c(k) 
    &=& \frac{1}{\sqrt{\beta N}}
        \int_{0}^{\beta} {\rm d}\tau \sum_{i} 
        c_i(\tau)
        e^{i\omega_{n}\tau}
        e^{i{\bf k}\cdot{\bf r}_i},\\
\Delta_{\alpha}(q) 
    &=& \frac{1}{\sqrt{\beta N}}
        \int_{0}^{\beta} {\rm d}\tau \sum_{i} 
        \Delta_{\alpha}({\bf r}_i,\tau)
        e^{i\omega_{\nu}\tau}
        e^{i{\bf q}\cdot{\bf r}_i},
\end{eqnarray}
in which $k=(i\omega_n,{\bf k})$ [$q=(i\omega_{\nu},{\bf q})$] where the
fermion [boson] Matsubara frequencies are $\omega_n =
(2n+1)\frac{\pi}{\beta}$ [$\omega_{\nu}= (2\nu)\frac{\pi}{\beta}$].  

It is then a straightforward matter to integrate out the vestigial
fermionic degrees of freedom.  The purely bosonic action then reads
\begin{equation}
{\cal S} = {\cal S}_0 + {\cal S}_b,
\end{equation}
where the non-interacting and interacting terms are respectively,
\begin{eqnarray}
{\cal S}_0 &=& -{\rm Tr}\ln({\bf G}_0),\\
{\cal S}_b &=& -\frac{V}{2}\sum_{q,\alpha}|\Delta_{\alpha}(q)|^2 + {\rm
Tr}\ln({\bf 1}-{\bf V}{\bf G}_0).
\end{eqnarray}
The first term, ${\cal S}_0$, corresponding to the free electron action
depends only on the Nambu Green's function matrix,
\begin{eqnarray}
{\bf G}_0(k,k^{\prime}) = 
   \left(\begin{array}{cc}
        G_0(k) & 0\\
        0 & -G_0^*(k)
   \end{array}\right)\delta_{kk^{\prime}},
\end{eqnarray}
in which $G_0(k)=(i\omega_n-\xi_{\bf k})^{-1}$ is the free fermion
propagator, and $\xi_{\bf k}=\epsilon_{\bf k}-\mu$.  For nearest
neighbour hopping the band-energy is $\epsilon_{\bf k}=-2t\sum_{i=x,y}
\cos(k_i)$.

The second term, ${\cal S}_b$, which describes the effects of
electron-electron interaction, depends not only on the electron Green's
function but also on the bosonic off-diagonal potential matrix,
\begin{eqnarray}
{\bf V}(k,k^{\prime}) = 
    \frac{V}{\sqrt{2\beta N}}
        \left(
        \begin{array}{cc}
            0 & \Delta_{\alpha}(k-k^{\prime})\\
            \Delta^*_{\alpha}(-k+k^{\prime}) & 0
        \end{array}
        \right)\zeta^{\alpha}_{{\bf k},{\bf k}^{\prime}},
\end{eqnarray}
where the Einstein summation convention has been used for repeated
Greek indices and,
\begin{eqnarray}
\zeta^{\alpha}_{{\bf k},{\bf k}^{\prime}} =
        \frac{1}{2}\Big(e^{ik_x}+e^{-ik^{\prime}_x}\Big)
    \pm \frac{1}{2}\Big(e^{ik_y}+e^{-ik^{\prime}_y}\Big),\quad \alpha = s,d.
\end{eqnarray}
In the limit that ${\bf k}={\bf k}^{\prime}$ these become,
\begin{equation}
\zeta^{\alpha}({\bf k})
= \zeta^{\alpha}_{{\bf k},{\bf k}}
=   \left\{\begin{array}{cl}
        \cos(k_x)+\cos(k_y) & \alpha = s,\\
        \cos(k_x)-\cos(k_y) & \alpha = d.
    \end{array}\right.
\end{equation}
which reflect the point group symmetries of the order
parameters.

\begin{figure}
\centerline{\epsfig{file=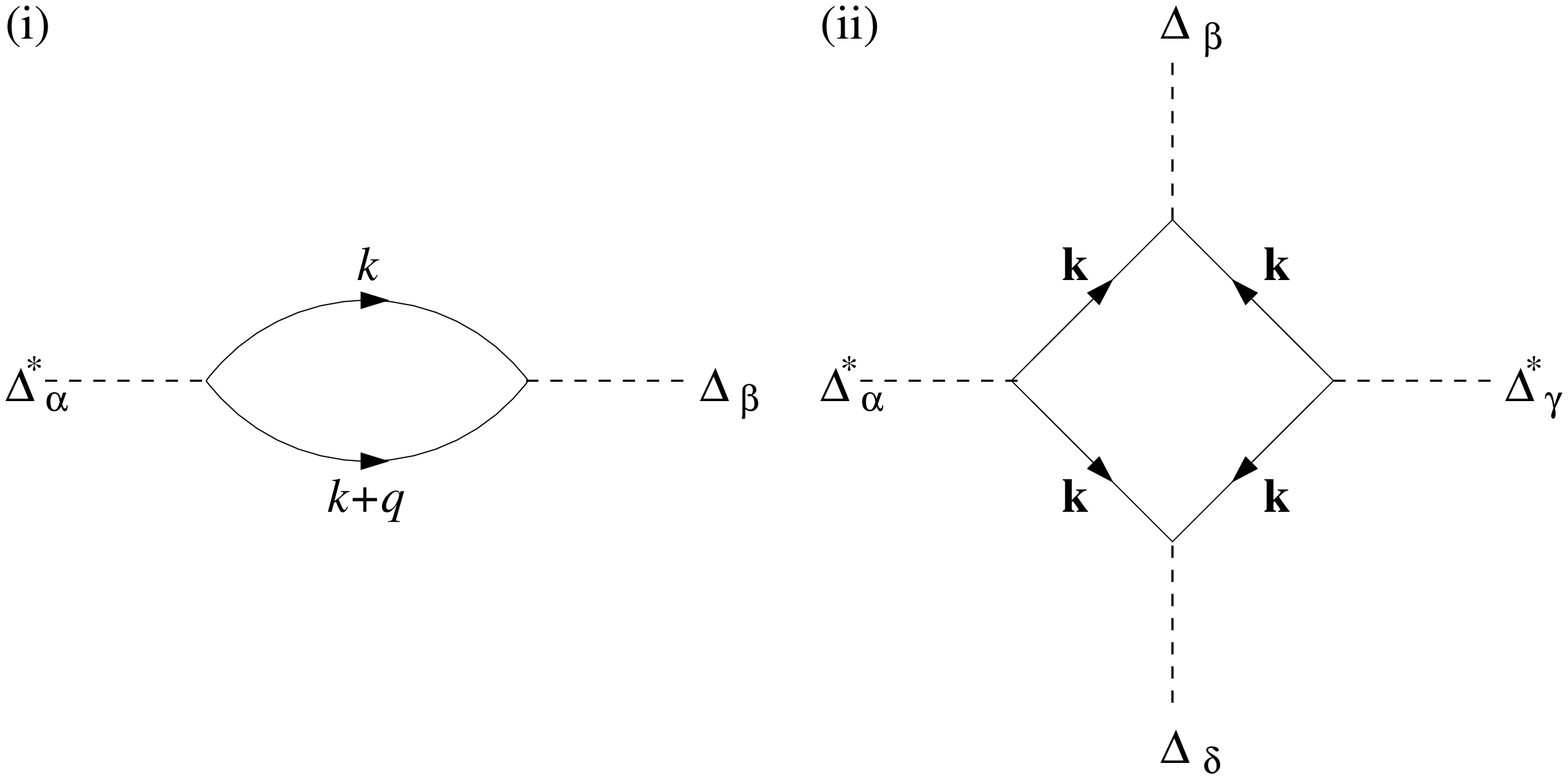,width=11.0cm,angle=0}}
\caption{(i) Quadratic and (ii) quartic terms in the expansion of ${\rm
Tr} \ln \left( {\bf 1} - {\bf V}{\bf G}_0 \right)$.}
\label{feyn_quartic_sd}
\end{figure}

Expanding the logarithm in ${\cal S}_b$ as a power series to fourth order,
only the second and fourth order terms are non-zero.  They are shown
diagrammatically in Fig. \ref{feyn_quartic_sd}

After a trivial rescaling of $\Delta$ by $V^{\frac{1}{2}}$, the bosonic
action to second order reads,
\begin{eqnarray}
{\cal S}_b^{(2)} = 
    -\frac{1}{2}\sum_{q}
      \left(
         \delta^{\alpha\beta}-V\chi^{\alpha\beta}_0(q)
      \right)
      \Delta^*_{\alpha}(q)\Delta_{\beta}(q),
\end{eqnarray}
where the particle-particle susceptibilities are,
\begin{eqnarray}
\label{sd_X}
\chi^{\alpha\beta}_0(q)
 &=& +\frac{1}{\beta N}\sum_{k}G_0(k)G_0^*(k+q)
    \zeta^{\alpha}_{{\bf k},{\bf k}+{\bf q}}
    \zeta^{\beta}_{{\bf k}+{\bf q},{\bf k}},\\
 &=& -\frac{1}{N}\sum_{{\bf k}}
   \left[
   \frac{n_f(\xi_{{\bf k}+\frac{\bf q}{2}})
        +n_f(\xi_{{\bf k}-\frac{\bf q}{2}})-1}
        {i \omega_{\nu}+\xi_{{\bf k}+\frac{\bf q}{2}}
                       +\xi_{{\bf k}-\frac{\bf q}{2}}}
   \right]
    \zeta^{\alpha}_{{\bf k}-\frac{\bf q}{2},{\bf k}+\frac{\bf q}{2}}
    \zeta^{\beta}_{{\bf k}+\frac{\bf q}{2},{\bf k}-\frac{\bf q}{2}}
,\nonumber
\end{eqnarray}
where $n_f(x)\equiv[e^{\beta x}+1]^{-1}$ is the Fermi function.

Regardless of the flavour indices, $\chi^{\alpha\beta}_0(q)$ is a maximum
at zero frequency and zero momentum.  Keeping only relevant terms,
expansion of the Gaussian action about $q=0$ gives,
\begin{equation}
\label{sd_Gaussian_action}
{\cal S}_b^{(2)} = 
-\frac{1}{2} \sum_{q}
      \left(
        r^{\alpha\beta}
        +\frac{q_i^2}{2m^i_{\alpha\beta}}
        -id^{\alpha\beta}\omega_{\nu}
      \right)
      \Delta^*_{\alpha}(q) \Delta^{}_{\beta}(q).
\end{equation}
The frequency dependence has been kept, to lowest order in $\omega_{\nu}$,
in order to look at the effects of quantum fluctuations.

The constant term is given by,
\begin{equation}
r^{\alpha\beta} =
      \left(
         \delta^{\alpha\beta}-\frac{V}{2}\sum_{\epsilon}
         \frac{\tanh\left(\beta\xi/2\right)}{\xi}
         N^{\alpha\beta}(\epsilon)
      \right),
\end{equation}
where the weighted densities of states (WDOS) are
\begin{equation}
N^{\alpha\beta}(\epsilon)
    =\frac{1}{N}\sum_{{\bf k}}
        \zeta^{\alpha}({\bf k})\zeta^{\beta}({\bf k})
        \delta(\epsilon - \epsilon_{{\bf k}}).
\end{equation}
For the nearest neighbour hopping square lattice the WDOS are plotted in
Fig. \ref{sd_WDOS}.  As one can see from the figure the $d$-wave WDOS is
strongly peaked around the normal state van Hove peak at half-filling,
whilst the (extended) $s$-wave WDOS is suppressed at that point but large
near the band edges.

Note that the square symmetry of the lattice forbids the existence of cross-terms
such as $r^{sd}$ so that $N^{sd}$ is zero.  Whilst $\zeta^{s}({\bf k})$ is
even under the exchange of $x$- and $y$-momenta, $\zeta^{d}({\bf k})$ is
odd under this operation.  The products $\zeta^{s}\zeta^{s}$ and
$\zeta^{d}\zeta^{d}$ are both even but $\zeta^{s}\zeta^{d}$ is odd.
Provided $\epsilon_{{\bf k}}$ is even under $x \leftrightarrow y$, $N^{sd}$
must be zero.  However, if the $x \leftrightarrow y$ symmetry in
$\epsilon_{{\bf k}}$ is broken, say by an orthorhombic distortion, then
$r^{sd}$ need not be zero.

The masses come from the second derivatives (with respect to momentum) of
the static susceptibilty,
$\chi^{\alpha\beta}_0({\bf q})=\chi^{\alpha\beta}_0(i\omega_{\nu}=0,{\bf
q})$ at ${\bf q}=0$.  Generally, for any given band structure, $\epsilon_{\bf k}$,
they are,
\begin{eqnarray}
\frac{1}{2m^{ij}_{\alpha\beta}} 
&=& \frac{V}{8N} \sum_{\bf k}
    \Bigg[
        \frac{1}{m^*_{ij}({\bf k})}
        \left(
            \frac{n_f^{\prime}(\xi_{\bf k})}{\xi_{\bf k}}
            -\frac{n_f(\xi_{\bf k})-\frac{1}{2}}{\xi^2_{\bf k}}
        \right) 
        \zeta^{\alpha}({\bf k})\zeta^{\beta}({\bf k}) \\
&&\hspace{4cm}
       +\;v_i({\bf k}) v_j({\bf k})
        \left(
            \frac{n_f^{\prime\prime}(\xi_{\bf k})}{\xi_{\bf k}}
        \right) 
        \zeta^{\alpha}({\bf k})\zeta^{\beta}({\bf k})
    \Bigg],\nonumber
\end{eqnarray}
where a prime indicates differentiation and the band masses and
quasiparticle velocities are,
\begin{equation}
\frac{1}{m^*_{ij}({\bf k})} 
    =   \left( 
            \frac{ \partial^2 \epsilon_{\bf k}}{\partial k_i \partial k_j}
        \right),\qquad
v_i({\bf k}) 
    =  \left( 
            \frac{ \partial \epsilon_{\bf k}}{\partial k_i}
        \right).
\end{equation}
For the nearest neighbour square lattice only the diagonal terms,
$m^*_{ii}({\bf k})$ and $v_i^2({\bf k})$, are non-zero and even:
\begin{equation}
\frac{1}{m^*_{ii}({\bf k})} 
    = 2t \cos(k_i),\qquad
v_i({\bf k})^2
    = 4t^2 \sin^2(k_i).
\end{equation}
Unlike $r^{\alpha\beta}$, the mass in Eq. (\ref{sd_Gaussian_action})
is not purely diagonal in the flavour indices.  In addition to the diagonal
terms there are gradient mixing terms \cite{FederKallin97,Joynt90},
\begin{equation}
\Delta^*_s(q) 
\left(
    \frac{q_x^2 - q_y^2}{2m^x_{sd}}
\right)
\Delta^{}_d(q) +
\Delta^*_d(q) 
\left(
    \frac{q_x^2 - q_y^2}{2m^x_{sd}}
\right)
\Delta^{}_s(q),
\end{equation}
where $m^x_{sd}=-m^y_{sd}$.  They are allowed because they have the
symmetry of the square.  Under an exchange $x \leftrightarrow y$ both the
gradients and the $d$-wave order parameter change sign, leaving the whole
expression unchanged.

For weak coupling, such terms have little effect except in the vicinity of
inhomogeneities such as impurities or interfaces where the magnitude of the
order parameter is expected to be diminished.  In these regions, a gradient
in the dominant order parameter would drive the subdominant order parameter
to be non-zero.  So, for instance, near the surface of a bulk $d$-wave
superconductor an induced $s$-wave component might appear.  When we
consider the strong coupling limit we will see the importance of these
gradient mixing terms even in the bulk.

We now have expressions for $r^{\alpha\beta}$ and $m^i_{\alpha\beta}$,
leaving only $d^{\alpha\beta}$ to be determined.  The derivation of
$d^{\alpha\beta}$ requires some care \cite{Randeria95}.  The Matsubara
frequencies are discrete, therefore to expand in them we must first make
the analytic continuation $\chi^{\alpha\beta}(i\omega_{\nu},{\bf q}=0)
\rightarrow \chi^{\alpha\beta}(\omega+i0^+,{\bf q}=0)$ and then expand in
$\omega$.  We find,
\begin{equation}
d^{\alpha\beta} =
\frac{V}{4}\sum_{\epsilon}
    \frac{\tanh\left(\beta\xi/2\right)}{\xi^2}
    N^{\alpha\beta}(\epsilon)
-\frac{i \pi\beta V}{8}N^{\alpha\beta}(\mu).
\end{equation}
Like $r^{\alpha\beta}$, the tensor $d^{\alpha\beta}$ is diagonal in the
flavour indices.  The real part of $d^{\alpha\beta}$ represents the free
propagation of pairs, whilst the imaginary part represents the decay of
pairs into the continuum of fermionic states.  The significance of the two
terms, and their domains of applicability will be discussed when we
consider the effects of Gaussian fluctuations in section \ref{sect_GFA}.

The quartic part of the action is shown in Fig.
\ref{feyn_quartic_sd}(ii).  We associate a susceptibility with the fermion
loop:
\begin{eqnarray}
\label{sd_quartic_X}
\chi_4^{\alpha\beta\gamma\delta} 
&=& -\frac{1}{\beta N} \sum_{\bf k}
        |G_0(i\omega_n,{\bf k})|^4
        \zeta^{\alpha}({\bf k})
        \zeta^{\beta}({\bf k})
        \zeta^{\gamma}({\bf k}) 
        \zeta^{\delta}({\bf k}), \\
&=& -\frac{1}{2N} \sum_{\bf k}
         \left(
            \frac{n_f^{\prime}(\xi_{\bf k})}{\xi_{\bf k}^2}
           -\frac{n_f(\xi_{\bf k})-\frac{1}{2}}{\xi_{\bf k}^3}
        \right)
        \zeta^{\alpha}({\bf k})
        \zeta^{\beta}({\bf k})
        \zeta^{\gamma}({\bf k}) 
        \zeta^{\delta}({\bf k}), \nonumber
\end{eqnarray}
where $\alpha,\beta,\gamma,\delta$ denote the flavour indices at each
vertex.  Only the diagrams with an even number of $s$ and $d$ indices
contribute.  Thus, including the rescaling of $\Delta$ by $V^{\frac{1}{2}}$, the
bosonic action to fourth order reads,
\begin{eqnarray}
{\cal S}^{(4)}
&=& -\frac{1}{4} {\rm Tr}\left({\bf V}{\bf G}_0\right)^4,\\
&=& -u_s \sum_{\{q\}} 
        \Delta^*_s(q_1) \Delta^{}_s(q_2) \Delta^*_s(q_3) \Delta^{}_s(q_4)
        \times \delta(q_1-q_2+q_3-q_4)\nonumber\\
&& \qquad -
        u_d \sum_{\{q\}}
        \Delta^*_d(q_1) \Delta^{}_d(q_2) \Delta^*_d(q_3) \Delta^{}_d(q_4)
        \times \delta(q_1-q_2+q_3-q_4)\nonumber\\
&& \qquad -
        2u_x \sum_{\{q\}}
        \Delta^*_s(q_1) \Delta^{}_s(q_2) \Delta^*_d(q_3) \Delta^{}_d(q_4)
        \times \delta(q_1-q_2+q_3-q_4)\nonumber\\
&& \qquad -
        \frac{u_x}{2} \sum_{\{q\}}
        \Delta^*_s(q_1) \Delta^{}_d(q_2) \Delta^*_s(q_3) \Delta^{}_d(q_4)
        \times \delta(q_1-q_2+q_3-q_4)\nonumber\\
&& \qquad -
        \frac{u_x}{2} \sum_{\{q\}}
        \Delta^*_d(q_1) \Delta^{}_s(q_2) \Delta^*_d(q_3) \Delta^{}_s(q_4)
        \times \delta(q_1-q_2+q_3-q_4)\nonumber.
\end{eqnarray}
The three coefficients are,
\begin{eqnarray}
u_s &=& -\frac{V^2}{8 \beta N} \chi_4^{ssss},\\
u_d &=& -\frac{V^2}{8 \beta N} \chi_4^{dddd},\\
u_x &=& -\frac{V^2}{4 \beta N} \chi_4^{ssdd},
\end{eqnarray}
where we note that $\chi_4^{ssdd}=\chi_4^{sdsd}$.

This completes our derivation of the Landau-Ginzburg-Wilson action.  In the
following sections we will examine the saddle point (mean field) and
Gaussian fluctuation approximations to this action in order to derive the
phase diagram of the model and to study the crossover to strong-coupling superconductivity. 


\section{Mean field (Landau) theory}
\label{sect_MFT}


In the limit that the superconducting fields are static and spatially
uniform, the action, ${\cal S}_b$, is related to the Landau free energy
for interacting $s$- and $d$-wave superconductivity:
\begin{eqnarray}
\beta{\cal F} = -{\cal S}_b^{mf}
&=&
\frac{1}{2} r_s |\Delta_s|^2
   + \frac{1}{2} r_d |\Delta_d|^2
   +u_s|\Delta_s|^4
   +u_d|\Delta_d|^4 \\
&\;&\qquad\qquad\qquad\qquad
   +2u_x|\Delta_s|^2|\Delta_d|^2
   +\frac{u_x}{2}\left(
        \Delta_s\Delta_s
        \Delta^*_d\Delta^*_d
        +c.c.\right).\nonumber
\end{eqnarray}
This form for the free energy is identical to that found by Feder and
Kallin \cite{FederKallin97} and Joynt \cite{Joynt90} who derived it from
symmetry principles.

Rationalising the two cross-terms into one by writing the order
parameters in terms of their phases and moduli, so that
\begin{equation}
2u_x|\Delta_s|^2|\Delta_d|^2
    +\frac{u_x}{2}\left(
                            \Delta_s\Delta_s
                            \Delta^*_d\Delta^*_d
                            +c.c.
                     \right)
=2u_x\left(1+\frac{1}{2}\cos(2\Delta\theta)\right)|\Delta_s|^2|\Delta_d|^2,
\end{equation}
we see that the free energy, ${\cal F}$, depends on the phase difference,
$\Delta\theta$, between $\Delta_s$ and $\Delta_d$.

\begin{figure}
\centerline{\epsfig{file=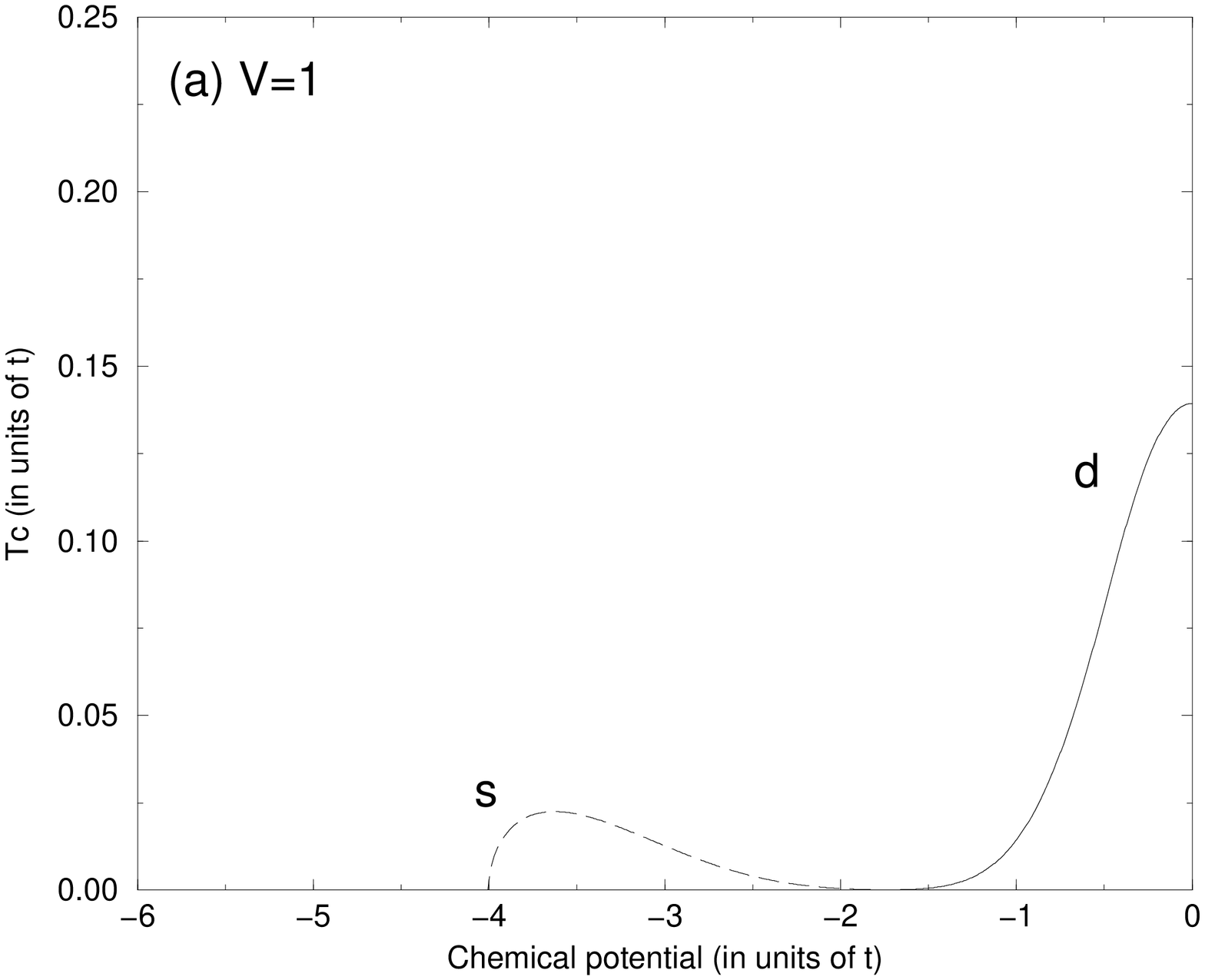,width=7.8cm,angle=-90}}
\centerline{\epsfig{file=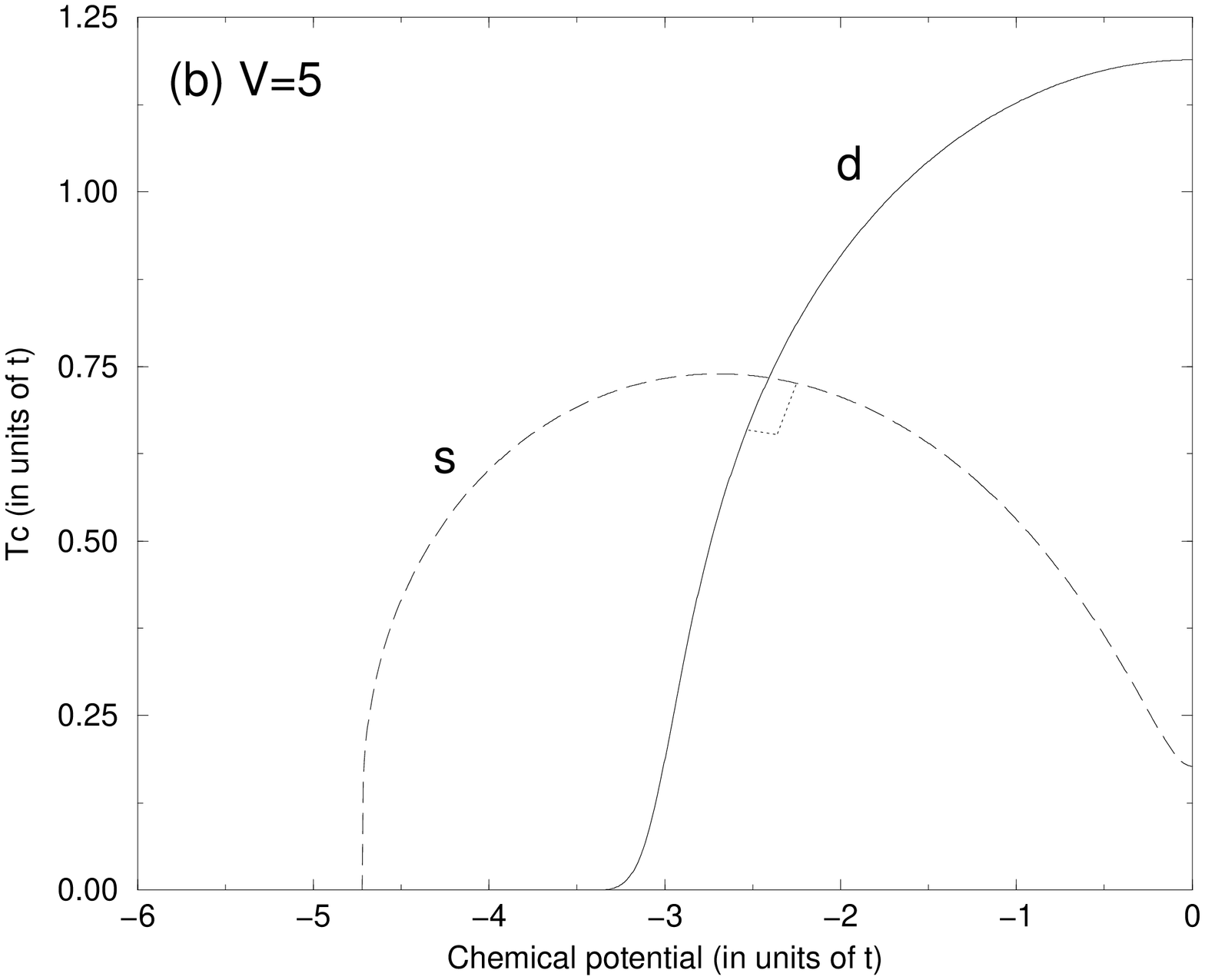,width=7.8cm,angle=-90}}
\centerline{\epsfig{file=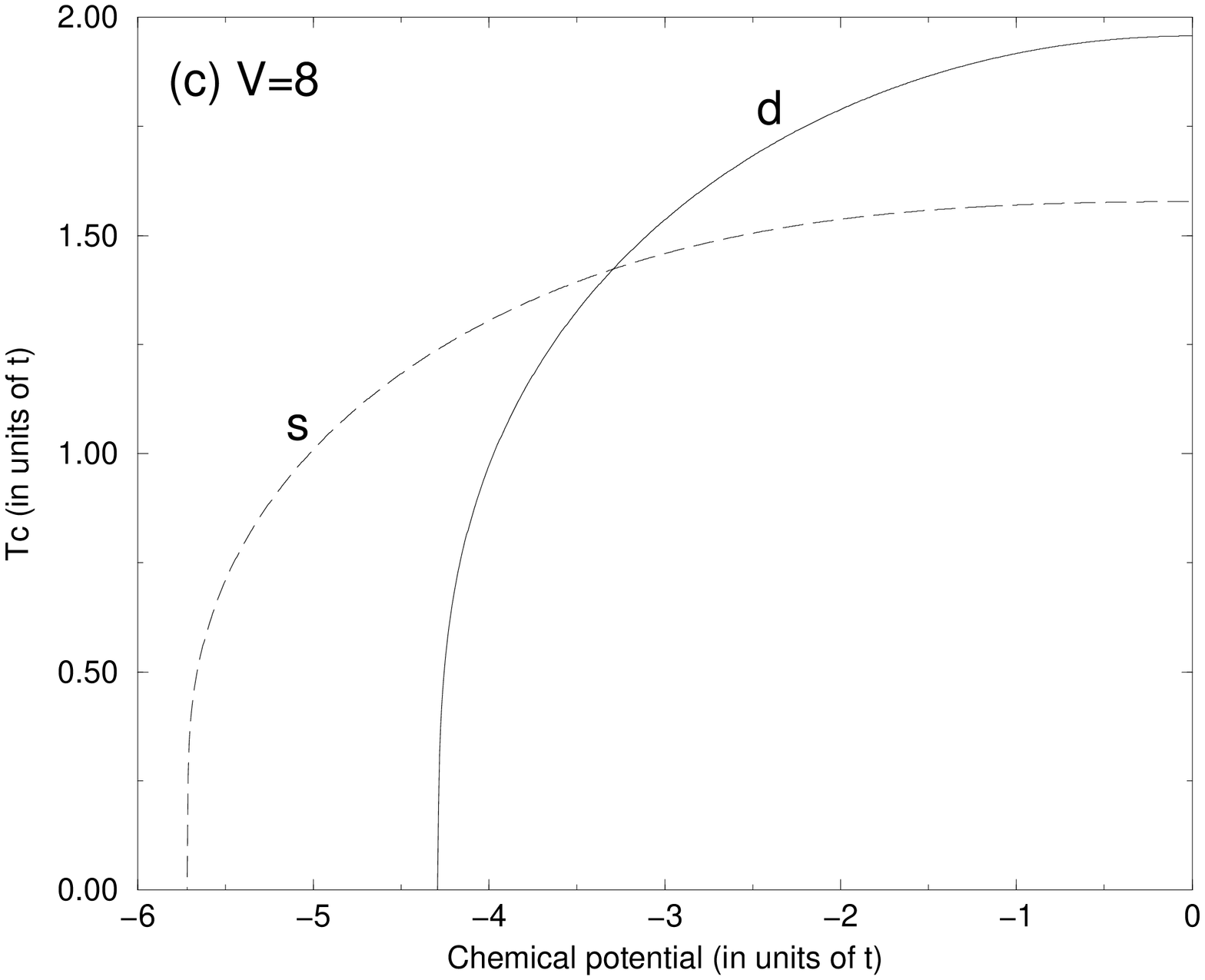,width=7.8cm,angle=-90}}
\caption{Mean field critical temperature, T$_c$, versus chemical potential,
$\mu$, for $s$- and $d$-wave superconductors with (a) $V=1t$, (b) $V=5t$
and (c) $V=8t$.}
\label{sd_crit_T}
\end{figure}

The saddle points of the free energy, ${\cal F}$, which for a uniform
system correspond to Hartree-Gor'kov mean field theory, are found by
simultaneously minimising ${\cal F}$ with respect to the magnitudes of the
fields and the phase difference, i.e.,
\begin{equation}
\label{sd_saddle}
\frac{\partial {\cal F}}{\partial |\Delta_s|}
    =\frac{\partial {\cal F}}{\partial |\Delta_d|}
    =\frac{\partial {\cal F}}{\partial \Delta\theta}=0.
\end{equation}
As a first approximation to solving these equations, we uncouple the two
order parameters and, for given $V$, solve for T$_c$ as a function of $\mu$
for each field independently of the other.  In this approximation, Eq.
(\ref{sd_saddle}) gives Stoner-like criteria for each of the critical
temperatures:
\begin{equation}
r_{\alpha}=0, \qquad \alpha=s,d.
\end{equation}
It is not possible to find analytic solutions to these equations except in
the large $V$ limit where we expect T$_c$ to be large too.  In this case we
find,
\begin{equation}
\lim_{V \rightarrow \infty} r_{\alpha}
= 1-\frac{V\beta}{4}\sum_{\epsilon}N^{\alpha\alpha}(\epsilon),
\end{equation}
which implies that, in this limit, the critical temperatures for
both $s$- and $d$-wave superconductivity are the same:
\begin{equation}
{\rm T}_c = \frac{V}{4}.
\end{equation}
Computationally, however, the solution of $r_{\alpha}=0$ is a trivial
matter for all $V$.  The results are shown in Figs.
\ref{sd_crit_T}(a)-(c) for three different values.  The maximum value of
the ordinate has been chosen as $V/4$ for each one so that the approach to
the strong-coupling limit can be followed.

For small coupling, as in Fig. \ref{sd_crit_T}(a), the $s$-wave solutions
have their highest T$_c$ at the edges of the band whilst the $d$-wave
solutions have theirs at the band centre.  These places correspond to where
the respective WDOS of each is the largest.  The effect of the van Hove
singularity in the centre of the $d$-wave WDOS is to produce a high
transition temperature in that channel, just as would be predicted by
conventional BCS theory in which it is the combination
$VN^{\alpha\alpha}(0)$ that determines T$_c$.  In fact, it has been
credibly suggested by some authors
\cite{Friedel89,NewnsTsueiPattnaik95,LMGAW98} that it is the van Hove
singularity, observed in most first-principles calculations
\cite{AndersenJL94,NovikovFreeman96}, combined with a conventional
weak-coupling electron-phonon or spin-fluctuation mechanism, that is
responsible for the high transition temperatures in the cuprate materials.
For the extended $s$-wave channel there is no van Hove singularity and the
transition temperatures are typically a factor of five lower than the
$d$-wave ones.  For small coupling, neither the $s$- nor the $d$-wave
critical temperatures are near the limiting value of $V/4$

As $V$ is increased, Fig. \ref{sd_crit_T}(b), the $s$-wave solutions
begin to occupy the centre of the band and the $d$-wave solutions spread
outwards towards the band edges.  The maximum T$_c$'s increasingly tend to
the strong-coupling value, $V/4$.  As $V$ increases further towards the
strong-coupling limit, Fig. \ref{sd_crit_T}(c), the $s$- and $d$-wave
solutions start to converge.  This is because the exact shape of the WDOS
becomes less important as more of it falls within $k_B$T$_c$ of the Fermi
surface.  Note that for large $V$ both the $s$- and $d$-wave solutions have
spread beyond the edges of the band.  This, and the crossover to local
pairs, will be discussed when we consider the effects of Gaussian
fluctuations in the next section.

Now that the phase diagram of the uncoupled order parameters is known, we
turn our attention to the coupled case.  From Figs. \ref{sd_crit_T}(b)
and (c) we see that there are regions where both the uncoupled $s$- and
$d$-wave order parameters are simultaneously below their critical
temperatures.  What is the nature of the pairing in these regions?
According to the values of the model parameters, two cases may be
distinguished: (i) For $u_s u_d <
u_x^2\left(1+\frac{1}{2}\cos(2\Delta\theta)\right)^2$ the pure $s$- and
$d$-wave phases are separated by a single first order transition which
meets the two second order metal-superconductor lines at a bicritical
point; (ii) For $u_s u_d >
u_x^2\left(1+\frac{1}{2}\cos(2\Delta\theta)\right)^2$ an intermediate
phase of mixed symmetry ($\Delta_s \neq 0, \Delta_d \neq 0$) is
permissible, separated from each of the two pure phases by a line of
continuous second order transitions.  These meet the metal-superconductor
lines at a tetracritical point.

In the mixed phase the value of the phase
difference, $\Delta\theta$, is of paramount importance.  It is easily found
from Eq. (\ref{sd_saddle}). Depending on the sign of $u_x$ it can
assume either of two values:
\begin{equation}
\Delta\theta
=   \left\{\begin{array}{clll}
        \pi/2 & u_x > 0 & \Rightarrow & s \pm id,\\
        0     & u_x < 0 & \Rightarrow & s \pm  d.
    \end{array}\right.
\end{equation}
The two solutions represent fundamentally different phases.  For instance,
the $s \pm d$ phase is invariant under time reversal, whilst the $s \pm id$
phase breaks this symmetry (time reversal involves complex conjugation).
Several authors have suggested that the existence of a time reversal
symmetry broken order parameter might be associated with spontaneous
generation of currents at twin boundaries \cite{BelzigBS98} or surfaces
\cite{FogelstromRS97}.  This effect has been observed experimentally as a
splitting of the zero bias conductance peak in zero magnetic field
\cite{Covington97}.  No such effect would be observed in the $s \pm d$ phase.

Assuming that we are below the critical temperatures of both the $s$- and
$d$-wave order parameters, all possible states of the model can be
parametrised by just two dimensionless variables,
\begin{eqnarray}
U_s &=& \frac{u_s}{u_d}\left|\frac{r_d}{r_s}\right|^2,\\
U_x &=& \frac{u_x\left(1+\frac{1}{2}\cos(2\Delta\theta)\right)}
             {u_d}\left|\frac{r_d}{r_s}\right|,
\end{eqnarray}
where we remember that $\Delta\theta$ depends on the sign of $U_x$.  

\begin{figure}
\centerline{\epsfig{file=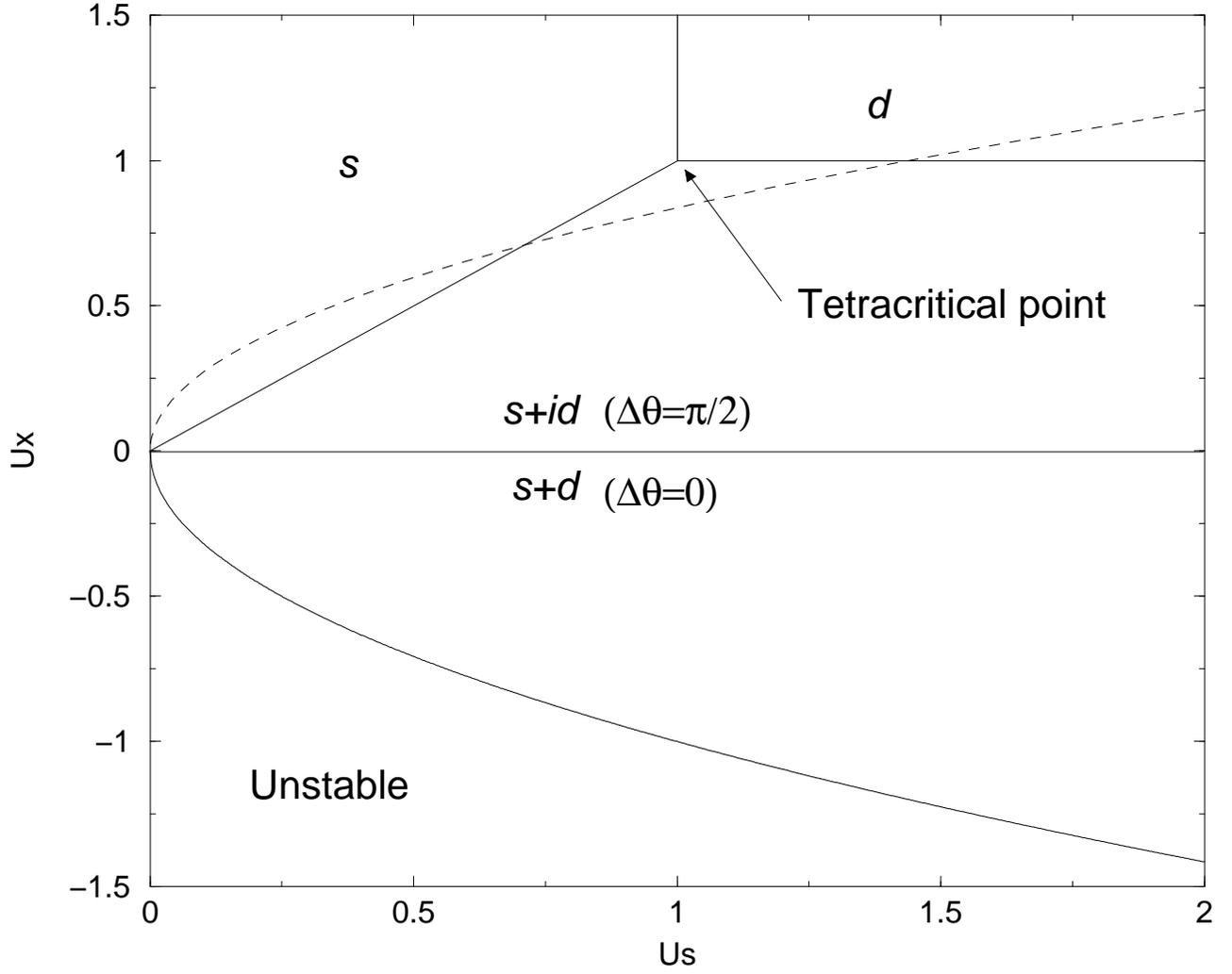,width=14.0cm,angle=-90}}
\caption{Plot of the stable phases of the Landau free energy as a function
of the two dimensionless parameters, $U_s$ and $U_x$.  The dashed line
shows parameters calculated from extended Hubbard model parameters near the
tetracritical point.}
\label{sd_landau_plot}
\end{figure}

As the values of these two parameters vary there are transitions between
the pure $s$- and $d$-wave states and the two different mixed phases
(separated by a line of first order transitions at $U_x=0$).  The stable
superconducting phases are shown in Fig. \ref{sd_landau_plot} as a
function of $U_s$ and $U_x$.  

For $V=5t$, the variables ($U_s,U_x$) were evaluated along a path in
($\mu$,T) space shown in Fig. \ref{sd_crit_T}(b) as a dotted line.  The
resulting values are plotted as the dashed trajectory in
\ref{sd_landau_plot}.  Starting in the pure $d$-wave sector at T = T$_c^s$,
we remain in the pure $d$ state even though we are below T$_c^s$: the
$s$-wave is suppressed by the dominant $d$-wave order parameter.
Eventually we cross the {\em second order} phase boundary ($d \rightarrow s
\pm id$) into the time reversal symmetry breaking mixed $s+id$ state.
With increasing $\mu$ the proportion of $s$-wave in the admixture rises and
the $d$-wave portion is diminished until the, now dominant, $s$-wave
totally supresses the $d$-wave and the second transition ($s \pm id
\rightarrow s$) occurs.  The trajectory terminates in the $s$-wave sector
when T=T$_c^d$.  We observe qualitatively identical behaviour for all $V$. 

The other, $s+d$, mixed phase does not occur spontaneously for any
parameters derived from the Hubbard model because of the form of the
quartic susceptibility, Eq. (\ref{sd_quartic_X}), which ensures that
$u_x$ is always positive.  To see an $s+d$ phase it is necessary to include
an orthorhombic distortion in the lattice, so that the hoppings in the $x$
and $y$ directions are different.  Orthorhombic anisotropy, as present in
YBCO, introduces an extra term into the free energy proportional to
$\Delta^*_s \Delta_d + \Delta^*_d \Delta_s$.  With a term like this a
finite value for one of the components drives the other to be non-zero too.
This effect has been studied elsewhere \cite{BelzigBS98} so we do not
consider it here.


\section{Gaussian fluctuation theory}
\label{sect_GFA}


It is clear from Figs. \ref{sd_crit_T}(a)-(c) that, as the coupling $V$
increases, T$_c$ starts to become finite even below the bottom (top) of the
electronic band at $\mu=-4t$ ($+4t$).  For such values of the chemical
potential, fermions can only exist as bound pairs, not as single particles.
A similar phenomenon has been noted for the local $s$-wave case in Jellium
by several authors
\cite{NSM85,DreschlerZwerger92,SadeMeloRanderiaEngelbrecht93,RanderiaDuanShieh90}
who concluded that, in the low density limit, Bose-Einstein condensation of
these pairs constitutes the mechanism for superconductivity.

\begin{figure}
\centerline{\epsfig{file=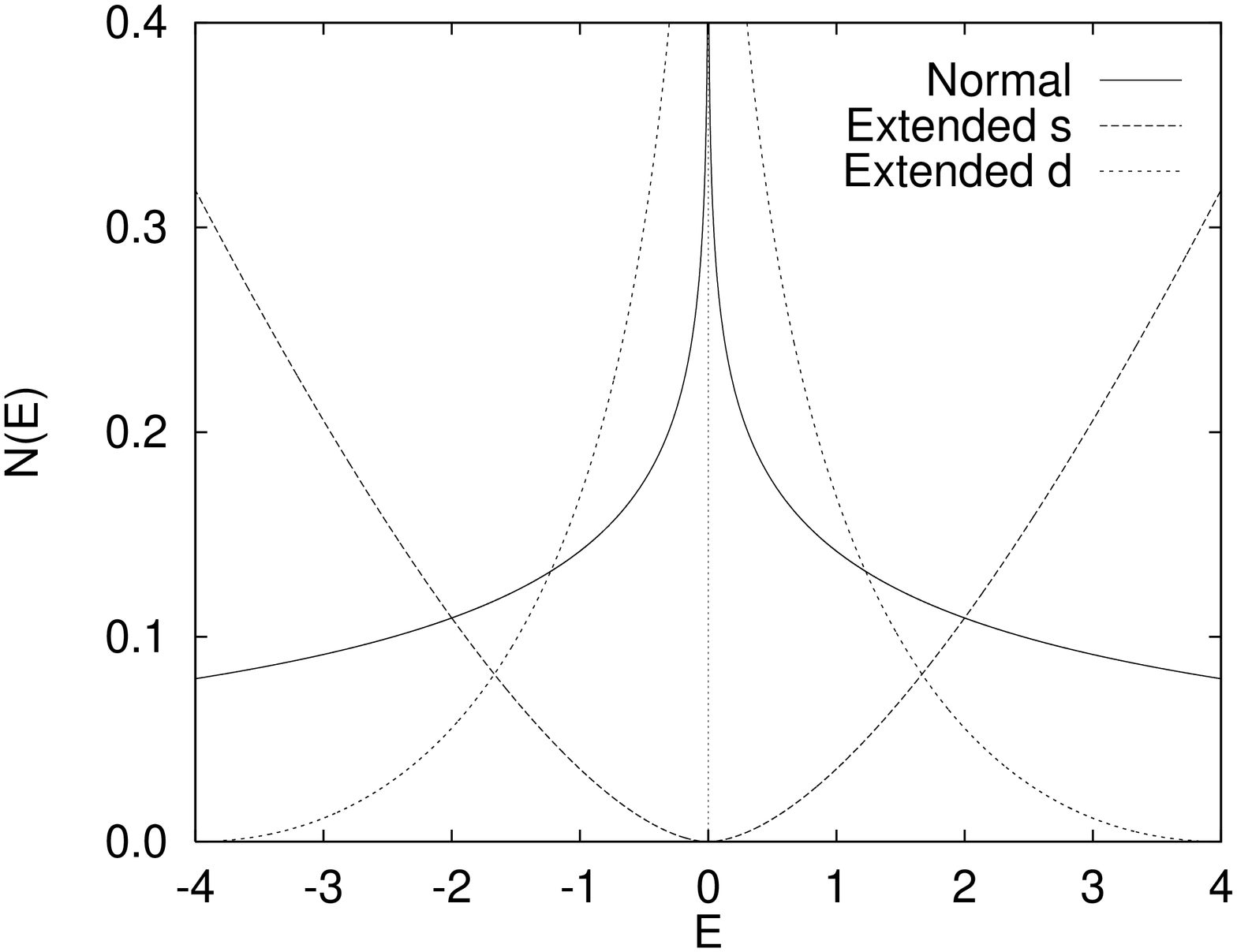,width=15.0cm,angle=-90}}
\caption{Weighted densities of states (WDOS) above T$_c$ for the normal
state and the extended $s$- and $d$-wave states.}
\label{sd_WDOS}
\end{figure}

\begin{figure}
\centerline{\epsfig{file=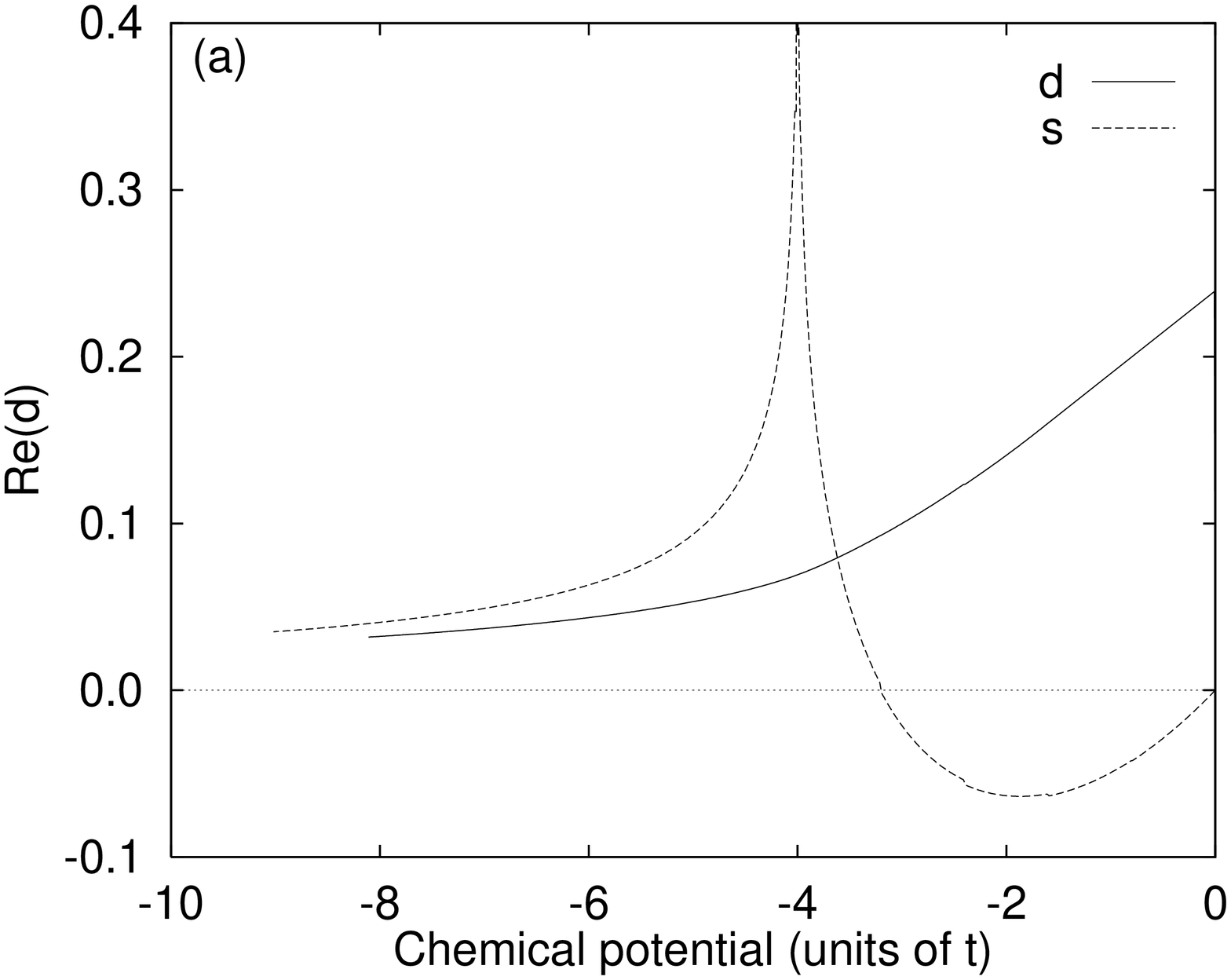,width=8.0cm,angle=-90}}
\centerline{\epsfig{file=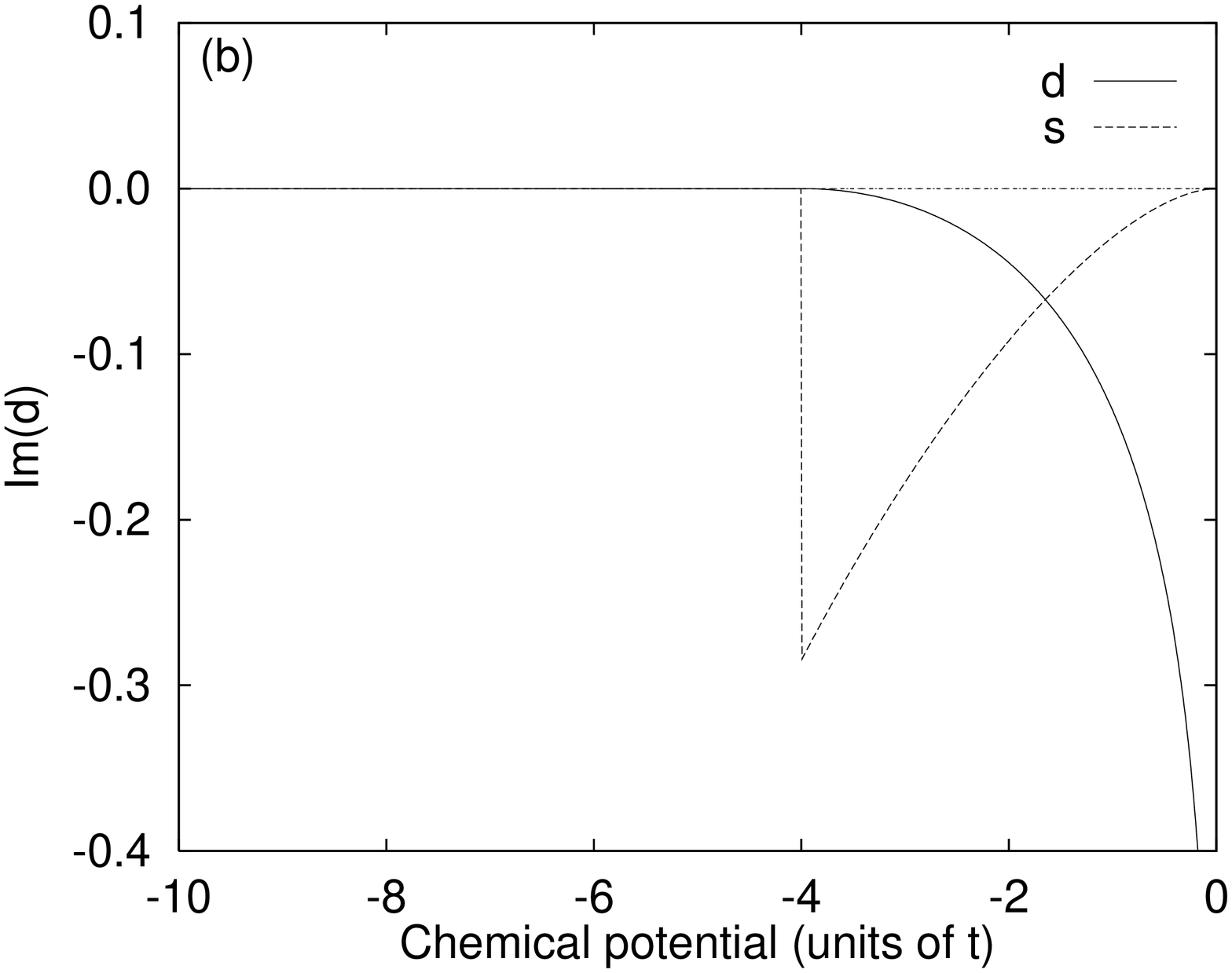,width=8.0cm,angle=-90}}
\caption{(a) Real and (b) imaginary parts of the frequency
coefficient, $d$(T$_c$,$\mu$), for extended $s$- and $d$-wave
superconductivity at $V=16t$.}
\label{sd_red_imd}
\end{figure}

As $V$ increases, the mean field approximation becomes increasingly
poor and the effects of fluctuations must be included.  Specifically we want to
know how T$_c$, as a function of filling, is affected.  As it stands, the
functional, ${\cal S}_b$, is too complicated.  We therefore employ the
Gaussian approximation in which terms of ${\cal O}(|\Delta|^4)$ and above are
neglected.  The free energy can be evaluated exactly in this
approximation.  The action is,
\begin{equation}
\label{sd_gaussian_action}
{\cal S}_b = 
-\frac{1}{2} \sum_{q}
      \left(
        r^{\alpha\beta}
        +\frac{q_i^2}{2m^i_{\alpha\beta}}
        -id^{\alpha\beta}\omega_{\nu}
      \right)
      \Delta^*_{\alpha}(q) \Delta^{}_{\beta}(q).
\end{equation}
There are two important limits to consider: weak coupling/high density
($|\mu|<4t$) and strong coupling/low density ($|\mu|>4t$).  Weak coupling
corresponds to BCS-like Cooper-pair superconductivity.  In this limit the
imaginary part of $d$ is dominant so the dynamics are relaxational: pairs
have a finite lifetime after which they decay into the continuum of
fermionic states.  As the coupling strength is increased, or the density is
lowered, the chemical potential can fall below the bottom of the band.  In
this case the imaginary part of $d$ is zero, see Fig. \ref{sd_red_imd},
so the pairs become infinitely long lived and the physics is that of
propagating pairs.  This is the Bose limit.  The manner in which the
crossover comes about is quite different for the $s$- and $d$-wave
superconductors because of form of the WDOS is different for each.  The
$s$-wave WDOS (see Fig. \ref{sd_WDOS}) shows a sharp band-edge so that
the transition from just inside to just outside the band is sharp too,
whilst the $d$-wave WDOS (see Fig. \ref{sd_WDOS}) goes smoothly to zero
so the transition is smooth.

Although our theory provides an interpolation scheme between the strongly
and weakly interacting regimes, its validity in the crossover region, near
$|\mu|=4$, is doubtful.  In this regime, fluctuation corrections
\cite{Haussmann93} of order $k_F a_F$ (where $k_F$ is the
Fermi vector and $a_F$ is the fermion scattering length), which are
vanishingly small in the extreme limits, become large signaling a breakdown
of the theory.  In the preceeding sections we studied the weak limit in
some detail.  In this section we will go on to examine the strong coupling
limit, omitting the intermediate regime.  This is regrettable, especially
as it is just this regime that is most relevant to the cuprates, but
unavoidable within the current formulation.


\subsection{On-site (local) pairing}
\label{subsect_GFA_onsite}


As a test of the performance of the Gaussian theory we will first consider
the on-site, local $s$-wave superconducting Hubbard model.  The Gaussian
action for it is, 
\begin{equation}
{\cal S}_b = 
-\frac{1}{2} \sum_{q}
      \left(
        r
        +\frac{{\bf q} \cdot {\bf q}}{2m}
        -id\omega_{\nu}
      \right)
      |\Delta_q|^2,
\end{equation}
where $\Delta_q$ is the local $s$-wave field and the parameters, $r,m$ and
$d$, are found from those of the nearest neighbour extended Hubbard model
by replacing $V$ with $U$ and setting $\zeta^{{\rm local}\;s}({\bf k})=1$.
Scaling the fields, we can set the frequency coefficient to unity, in which
case the action resembles that of free bosons:
\begin{equation}
{\cal S}_b = 
-\frac{1}{2} \sum_{q}
      \left(
        -\mu_b
        +\frac{{\bf q} \cdot {\bf q}}{2M}
        -i\omega_{\nu}
      \right)
      |\Delta_q|^2,
\end{equation}
with $\mu_b=-r/d$ the bosonic chemical potential, the zero of which signals
the onset of Bose-Einstein condensation, and $M=md$ the boson mass.  It is
well known \cite{Fradkin91} that as the on-site coupling, $U$, increases,
so too does the mass of the bosons.  This can be understood by considering
the mechanism of the bosonic hopping. The hopping of a pair consists of
three stages: (i) A pair occupies a single site. (ii) {\em Virtual
ionisation} - one electron hops onto a neighbouring site, gaining a kinetic
energy, $t$, but sacrificing the pair-binding energy, $U$. (iii) {\em Pair
recombination} - the second electron hops onto the neighbouring site too,
gaining a kinetic energy, $t$, and pair-binding energy, $U$.  The matrix
element for the whole hopping process, proportional to one over the boson
mass, is $t^2/U$.  We now show that our Gaussian theory can produce this
effect.  

In the extreme large $U$ limit, where the density is very
low and $U$/T$_c \gg 1$, the electronic chemical potential equals half the
binding energy of a Bose pair:
\begin{equation}
\lim_{U \rightarrow \infty} \mu = -\frac{U}{2}.
\end{equation}
In the same limit we find the frequency coefficient,
\begin{equation}
\lim_{U \rightarrow \infty} d
= \frac{U}{4}\sum_{\epsilon}
    \frac{1}{\xi^2}
    N(\epsilon)
= \frac{U}{4\mu^2}
= \frac{1}{U},
\end{equation}
and the unscaled mass,
\begin{eqnarray}
\lim_{U \rightarrow \infty} \frac{1}{2m^i} 
&=& \frac{U}{16N} \sum_{\bf k}
        \frac{1}{m^*_{ii}({\bf k})}
            \frac{1}{\xi^2_{\bf k}}, \\
&=& \frac{U}{16N} \sum_{\bf k}
        \frac{1}{m^*_{ii}({\bf k})}
        \left(  
            \frac{1}{\mu^2}
           +\frac{2\epsilon_{\bf k}}{\mu^3}
           + \ldots 
        \right), \\
&=& \frac{2t^2}{U^2},
\end{eqnarray}
where the first term in the expansion of $1/\xi^2_{\bf k}$ integrated to
zero.  Putting these together gives the asymptotic behaviour of the boson mass:
\begin{equation}
M = \frac{U}{4t^2},
\end{equation}
which is exactly as predicted by the virtual ionisation arguments.

We now evaluate the free energy. With the action in Gaussian form the
functional integral can be performed exactly to give,
\begin{eqnarray}
\beta{\cal F}_b = 
    -\ln \int {\cal D}\big[\Delta^*,\Delta] \; e^{{\cal S}_b}
&=& -\ln{\rm Det}^{-1}
        \left(
            -\mu_b + \frac{{\bf q}\cdot{\bf q}}{2M} - i\omega_{\nu}
        \right),\\
&=& {\rm Tr}\ln
        \left(
            -\mu_b + \frac{{\bf q}\cdot{\bf q}}{2M} - i\omega_{\nu}
        \right).
\end{eqnarray}
For given $\mu$ and $U$ Gaussian fluctuations do not alter T$_c$, but this
is not to say that they do not have any effect at all.  Following Randeria
{\em et al.} \cite{RanderiaDuanShieh90}, who examined an $s$-wave continuum
model, we calculate the fluctuation contribution to the fermion number
density, $n(\mu,T_c)$.  For fixed $n$, the changes in $\mu(n,{\rm T}_c)$
effectively renormalises T$_c(n)$.

The fermion number density, $n$, is derived from the free energy,
${\cal F}={\cal F}_0+{\cal F}_b$, by differentiation with respect to
the chemical potential, $\mu$:
\begin{equation}
\label{number}
n = n_0 + \delta n = -\frac{\partial{\cal F}}{\partial \mu}.
\end{equation}
The contribution from the free fermion action, ${\cal F}_0$, is
\begin{equation}
n_0 = 1 - \sum_{\epsilon}\tanh\left(\beta\xi/2\right)N(\epsilon).
\end{equation}
In the BCS regime, $|\mu|<4$, $\delta n$ is negligible and the mean field
results are unaltered.  As we approach the band edge, $|\mu|=4$, where our
theory is not strictly valid, the fluctuation contribution starts to become
important.  In the Bose limit, $|\mu|>4$, the fluctuation contribution to
$n$ is considerable.  In fact, in the $U \rightarrow \infty$ limit
$n(\mu,$T$_c) = \delta n(\mu,$T$_c)$.  The boson contribution is,
\begin{eqnarray}
\delta n    
&=&     \frac{1}{\beta}
        \frac{\partial \mu_b}{\partial \mu}
        \sum_{{\bf q}}\sum_{\nu}
        \frac{1}{-\mu_b+{\bf q}\cdot{\bf q}/{2M}-i\omega_{\nu}},\\
\label{boseint}
&=&     \frac{\partial \mu_b}{\partial \mu}
        \sum_{{\bf q}}
        n_{b}\left(-\mu_b+{\bf q}\cdot{\bf q}/2M\right),  
\end{eqnarray}
where the Bose occupation number is $n_{b}(x) \equiv [e^{\beta x}-1]^{-1}$.
For simplicity we have ignored small contributions from terms proportional
to the derivatives of $d$ and $m$ which tend to zero as $U$ increases.

The partial derivative of $r$ is,
\begin{equation}
\frac{\partial r}{\partial \mu} =
    -\frac{U}{2}\sum_{\epsilon}
    \left(
        \frac{\tanh\left(\beta\xi/2\right)}{\xi^2}
        +\frac{2n^{\prime}(\xi)}{\xi}
    \right)
    N(\epsilon), 
\end{equation}
which, for T=T$_c$, leads to the useful result that,
\begin{equation}
\lim_{U \rightarrow \infty}
\frac{\partial \mu_b}{\partial \mu} = 2.
\end{equation}
Combining this result with Eq. (\ref{boseint}) tells us that, in the low
density, strong interaction limit, the fermion density at the
critical temperature is,
\begin{equation}
\label{boseint2}
n(\mu,{\rm T}_c) = 2 \sum_{{\bf q}}
    n_{b}\left({\bf q}\cdot{\bf q}/2M\right),  
\end{equation}
i.e., all the fermions are bound into Bose pairs of mass $M$.

Unfortunately, there is a fly in the ointment: in two dimensions the
integral in Eq. (\ref{boseint2}) has an infrared logarithmic divergence
due to soft ${\bf q}=0$ modes .  This problem can be circumvented
be remembering that the cuprate materials are not really two dimensional;
rather they are highly anisotropic 3D materials.  To take account of this
we make the substitution
\begin{equation}
\frac{{\bf q}\cdot{\bf q}}{2M}\equiv
\frac{q_x^2}{2M}+\frac{q_y^2}{2M}\longrightarrow
\frac{q_x^2}{2M}+\frac{q_y^2}{2M}+a^2\frac{q_z^2}{2M},
\end{equation}
where $a^2 \equiv M/M^z \ll 1$, reflecting the low mobility of electrons
between CuO$_2$ planes.  With this substitution $n$ is always finite.
Assuming T$_c < \frac{a^2\pi^2}{2M}$ we can safely take the
limits of the ${\bf q}$ integral to infinity, giving,
\begin{equation}
n    
=   \frac{\zeta(3/2)}{a}
    \left(
        \frac{M{\rm T}_c}{\pi\sqrt[3]{2}}
    \right)^{\frac{3}{2}}\!\!\!,
\end{equation}
where $\zeta(\cdot)$ is the Riemann Zeta function.  Aside from the $1/a$
prefactor which accounts for the mass anisotropy, this is exactly the result we
would expect for a 3D gas of mass $M$ bosons with Bose-Einstein
condensation temperature T$_c$.  Rearranging this equation gives an
expression for T$_c$ at fixed density:
\begin{equation}
{\rm T}_c = 2.09 \frac{\left(an\right)^{2/3}}{M}.
\end{equation}
From this equation we can see the importance of the mass: a large mass
means a low T$_c$ and {\em vice versa}.  Thus, for the large $U$ limit in
which we are interested, increasing the coupling, which would ``normally''
(i.e., in mean field talk) increase the critical temperature, has the
paradoxical effect of lowering T$_c$ due to the localising effect it has on
the pairs. 

We expect to see analogous behaviour in our extended Hubbard model as we
increase the coupling, but instead of becoming localised on a site, our
bosons should become localised on bonds, becoming more and more massive,
i.e., less mobile, as $V$ increases.  


\subsection{Nearest neighbour (non-local) pairing}
\label{subsect_GFA_offsite}

\begin{figure}
\centerline{\epsfig{file=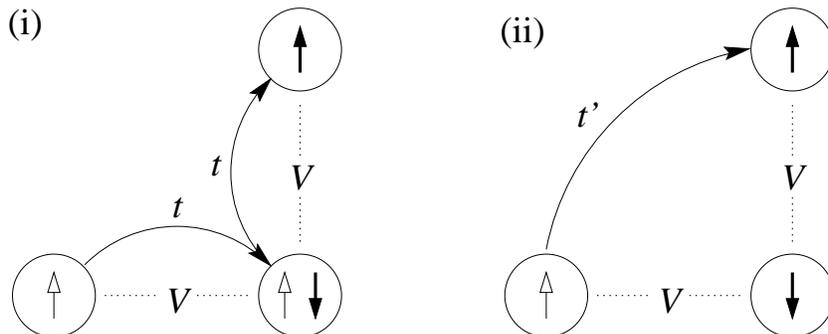,width=11cm,angle=0}}
\caption{Pair hopping in a tight binding model via (i) a two hops nearest
neighbour process, and (ii) a one hop next nearest neighbour process.}
\label{xy_hopping}
\end{figure}

The analysis for the extended Hubbard model proceeds along similar lines as
the on-site model but is complicated by the existence of two species of
bosons, $\Delta_s$ and $\Delta_d$, or, equivalently, $\Delta_x$ and
$\Delta_y$.  In the large $V$ limit we find,
\begin{eqnarray}
\lim_{V \rightarrow \infty} d^{\alpha\beta} 
&=& \frac{\delta_{\alpha\alpha}}{V},\\
\lim_{V \rightarrow \infty} \frac{1}{2m^i_{\alpha\beta}} 
&=& \frac{V}{16N} \sum_{\bf k}
        \frac{\zeta^{\alpha}({\bf k})\zeta^{\beta}({\bf k})}{m^*_{ii}({\bf k})}
            \frac{1}{\xi^2_{\bf k}}, \nonumber \\
&=& \frac{V}{16N} \sum_{\bf k}
        \frac{\zeta^{\alpha}({\bf k})\zeta^{\beta}({\bf k})}{m^*_{ii}({\bf k})}
        \left(  
            \frac{1}{\mu^2}
           +\frac{2\epsilon_{\bf k}}{\mu^3}
           + \ldots 
        \right).
\end{eqnarray}
For nearest neighbour hopping, these give the following bosonic masses:
\begin{equation}
(M^{x}_{\alpha\beta},M^{y}_{\alpha\beta})
=   \frac{V}{t^2} 
    \left\{
        \begin{array}{ll}
            (\frac{1}{9},\frac{1}{9}) & \alpha\beta=ss,\\
            (1,1) & \alpha\beta=dd,\\
            (1,-1) & \alpha\beta=sd.
        \end{array}
    \right.
\end{equation}
The anisotropy between the $s$ and $d$ boson masses arises because the form
factors for the $s$- and $d$-wave bosons, $\zeta^s\zeta^s$ and
$\zeta^d\zeta^d$, differ by the sign on $\cos(k_x)\cos(k_y)$. (This term does
not enter the $sd$ cross term.)  This is best understood in terms of the
$x$ and $y$ fields.  In that basis we see that the term comes from the
process, shown in Fig. \ref{xy_hopping}(i), by which an $x$ field becomes
a $y$ field.  Obviously such a process is sensitive to the relative phases
of the two bonds: zero phase difference is optimal for the hopping to occur
so the $s$-wave phase is preferred.

At this point, a pertinent question is, what happens if we change the bandstructure?  As an
example, consider adding a next nearest neighbour hopping, $t^{\prime}$, to
the Hamiltonian so that the band energy is,
\begin{equation}
\epsilon_{\bf k} = 
    -2t\left(\cos(k_x)+\cos(k_y)\right)
    -4t'\cos(k_x)\cos(k_y).
\end{equation}
The diagonal hopping allows the $x \leftrightarrow y$ process to occur
without a potential energy cost:  there is no intermediate state where $V$
is lost.  See Fig. \ref{xy_hopping}(ii)

In terms of the mathematics, the $t^{\prime}$ hopping means that the
integrand for the $1/\mu^2$ term in the expansion of $1/\xi_{\bf k}^2$ goes
like $\pm t^{\prime}\cos^2(k_x)\cos^2(k_y)$ which has the same sign over
the whole range of integration and is therefore finite (c.f. the nearest
neighbour hopping contribution at this order which has only odd powers of
$\cos(k_x)$ and $\cos(k_y)$ and therefore integrates to zero).  The masses
calculated with this term are:
\begin{equation}
\frac{1}{M_{\alpha\alpha}(t,t^{\prime})}
=   \left\{
        \begin{array}{ll}
            \frac{1}{V}(9t^2+t^{\prime}V) & \alpha = s,\\
            \frac{1}{V}(t^2-t^{\prime}V) & \alpha = d.
        \end{array}
    \right.
\end{equation}
If the product $t^{\prime}V$ is much smaller than $t^2$ the nearest
neighbour hopping can be safely ignored.  However, if this is not the case
then the corrections can be of great significance.  For instance, if
$t^{\prime}V \simeq t^2$ then the $d$-wave solution is unstable (has
negative mass), or if $t^{\prime}V \simeq -9t^2$ then the $s$-wave solution
is unstable.  In the limit $|t^{\prime}V| \gg t^2$ only one flavour of
boson is stable and its mass is {\em independent of} $V$:
\begin{equation}
\lim_{|t^{\prime}V| \rightarrow \infty} M(t,t^{\prime}) = \frac{1}{t^{\prime}}.
\end{equation}
Including a sufficiently large diagonal hopping, $t^{\prime}$, obviates the
need for pair hopping to occur via virtual ionisation.  Once this
requirement has been lifted, the lattice model behaves exactly as we would
expect a continuum model to behave (as shown in Fig. \ref{bcs_bose}).

Assuming for the moment that both boson modes are stable (i.e.,
$t^{\prime}=0$ or small) we resume our calculation.  The Gaussian free
energy is,
\begin{eqnarray}
\beta{\cal F}_b 
&=& {\rm Tr}\ln
        \left(
           -\mu_b^{\alpha\beta} + \frac{q_i^2}{2M^i_{\alpha\beta}} - i\omega_{\nu}
        \right),\\
&=& \sum_q \ln \left( \lambda_+ \lambda_- \right),
\end{eqnarray}
where we have used the cyclic nature of the trace to diagonalise the action
with respect to its flavour indices.  The eigenvalues, $\lambda_{\pm}$, are
the energies of the eigenmodes of the system which in general are linear
combinations of $s$ and $d$ (or $x$ and $y$) modes.  In the large $V$
limit, the eigenvalues are,
\begin{equation}
\lambda_{\pm} = -\frac{1}{2}\left(\mu^*_{ss}+\mu^*_{dd}\right) 
                +\frac{|{\bf q}|^2}{2M_{\pm}(\theta)}-i\omega_{\nu},
\end{equation}
where the masses, which depend on the angle $\theta =
\tan^{-1}(\frac{q_y}{q_x})$, are,
\begin{equation}
\frac{1}{M_{\pm}(\theta)}
= \frac{1}{M}
    \left(
        5 \pm \sqrt{17\cos^4(\theta)
                    +30\cos^2(\theta)\sin^2(\theta)
                    +17\sin^4(\theta)}
    \right),
\end{equation}
with $M=M_{dd}=V/t^2$ as a convenient mass unit.

\begin{figure}
\centerline{\epsfig{file=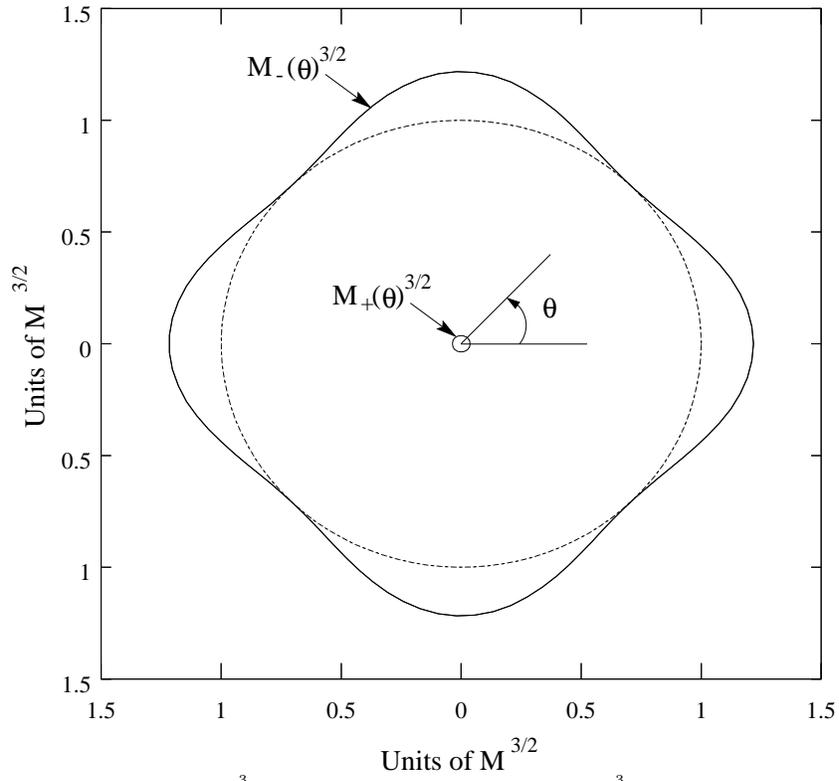,width=11cm,angle=0}}
\caption{Polar plot of $M_{\pm}(\theta)^{\frac{3}{2}} \sim
n_{\pm}(\theta)$. The unit circle is $M(\theta)^{\frac{3}{2}}$ for the pure
$d$-wave state.}
\label{mass_theta}
\end{figure}

With these definitions the bosonic number density is,
\begin{equation}
n(\mu,{\rm T}_c) 
= \frac{1}{2\pi}
    \int_{-\pi}^{+\pi} \!\! d\theta \;
        \Big\{
            n_{+}(\theta) + n_{-}(\theta)
        \Big\},
\end{equation}
where the densities of the two modes, as functions of angle
in ${\bf q}$-space, are
\begin{equation}
\label{bose_num_theta}
n_{\pm}(\theta)
=   \sqrt{\frac{2}{\pi}} 
    \frac{\zeta(3/2)}{a}
    \left(
        M_{\pm}(\theta){\rm T}_c
    \right)^{\frac{3}{2}}\!\!\!.
\end{equation}
We see from this that for given $n$ and T$_c$, the mode with the largest
mass will be the most populous.  This is graphically illustrated in Fig.
\ref{mass_theta} which shows the relative contributions of the two masses
to the angle-dependent density.

The flavour of boson with the smaller of the two masses, $M_{+}(\theta)$,
contributes negligibly ($\simeq 3\%$) to the total number density,
therefore the physics is determined by the boson with the largest mass,
$M_{-}(\theta)$.  From Fig. \ref{mass_theta} we see that the
$M_{-}(\theta)$ bosons are not distributed evenly throughout space - they
prefer to lie along one or other of the coordinate axes.  Relative to an
isotropic state, the bosons have been shifted away from the diagonals and
onto the axes.  For $M_{-}(\theta)$ as derived here, the effect is not very
large ($\simeq 10\%$ of the total), but by adjusting slightly the masses,
for instance by setting $M_{ss} \simeq M_{dd}$, nearly all the weight can
be shifted to the axes, leading to a very pronounced anisotropy.


\section{Discussion}


To understand how this might relate to the cuprates it is helpful to take a
step back.  In the derivations above we have assumed that all the electrons
are bound into Bose pairs, which is reasonable provided T$_c/V \ll 1$.
However, for higher temperatures, where T$/V \simeq 1$, this assumption is
not valid as thermal excitations lead to the break-up of pairs.  In the
extreme high temperature regime almost all the electrons are free.  Now,
starting in the high temperature regime, as the temperature is lowered, the
pairing fluctuations become more important and bosons start to form.  In
the Gaussian approximation, the onset of fluctuations is gradual, with a
maximum at T$_c$.  T$^*$ can be interpreted as the temperature where the
fluctuations become significant, which is not necessarily the same as
T$_c^{MF}$.  As the Bose number, $\delta n$, increases, the fermionic
spectral weight is correspondingly decreased according to Eq.
(\ref{number}).  Given the strongly anisotropic Bose number in Fig.
\ref{mass_theta} we can expect a similar anisotropy in the pseudogap.
Interestingly, the Bose number is maximum along the axes and minimum on the
diagonals which is just the same as the angle dependence of the pseudogap
observed by ARPES measurements.  Lowering the temperature further leads to
the creation of more bosons and the pseudogap becomes deeper.  Finally, at
the Bose-Einstein condensation temperature a real $d$-wave superconducting
gap will start to open, the maxima of which ``accidentally'' coincide with
the pseudogap regions where there are most bosons.

It is important here to make two comments.  Firstly, the pseudogap which we
observe is {\em not} a superconducting effect in the sense that its origin
does not lie in the spontaneous symmetry breaking of the superconducting
order parameter, but rather in the formation of two-particle states which
``eat'' the available single-particle spectral weight in certain
directions.  Secondly, although the angular variation of the pseudogap is
intimately related to the presence of $d$-wave pairs, the relation between
the two is not as simple as might first appear.  A single flavour order
parameter, either $s$-wave or $d$-wave, must have an {\em isotropic} mass
tensor (for $d$-wave this is shown as the dotted circle in Fig.
\ref{mass_theta}), otherwise the theory will not be invariant under the
square symmetry of the lattice.  The angular anisotropy of the pseudogap
comes not from the symmetry of a single order parameter, but from a
coupling of the $d$-wave to the $s$-wave bosons via the anisotropic
gradient, $\Delta^*_d(\nabla_x^2-\nabla_y^2)\Delta_s+c.c$., which has the
full square symmetry of the lattice.  Equivalently we could say that the
angular anisotropy is a reflection of the anisotropy in the masses of the
$x$ and $y$ bosons: an $x$ boson prefers to travel in the $x$ direction and
a $y$ boson prefers to travel in the $y$ direction.  Were the $x$ boson to
be totally confined to the $x$-axis, and the $y$ boson to the $y$-axis, in
which case $M_{ss} = M_{dd} = M$, the angular distribution of the bosons
would be sharply peaked along the axes and zero on the diagonals.

\begin{figure}
\centerline{\epsfig{file=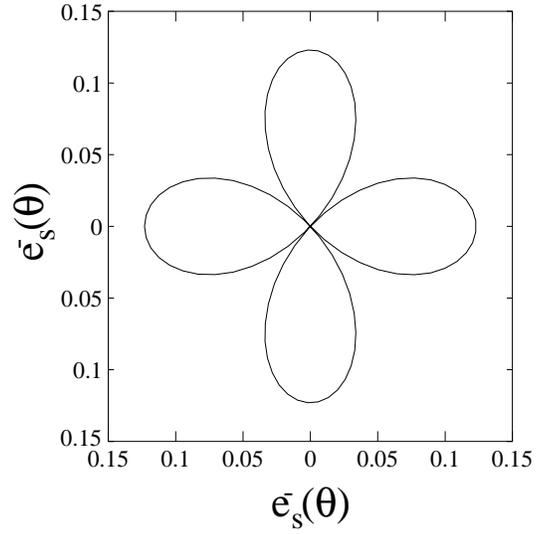,width=7cm,angle=0}}
\caption{Polar plot of the $s$-wave component, e$^-_s$, of the dominant
eigenvector.}
\label{eigenvector}
\end{figure}

We have seen that it is necessary to have both $s$- and $d$-wave order
parameters if the pseudogap is to have any angular dependence, so it is
natural to enquire about the nature of the superconducting state formed
below T$_c$.  In the preceeding calculations it has been assumed that in
the low density, large coupling limit, the critical temperatures for the
$s$- and $d$-wave fields become degenerate.  Now, just because two order
parameters have the same T$_c$ does not mean that they will be present in
the condensate in equal measure.  Ignoring the gradient coupling for the
moment, by analogy with Eq. (\ref{bose_num_theta}) it is easy to see
that, in a mixture of the two pure states, the heavier $d$ bosons will
dominate by a ratio of $M_s^{3/2}:M_d^{3/2}=1:27$.  Including the gradient
coupling terms the modes are admixtures of $s$ and $d$ bosons.  The
relative components can be found from the eigenvectors of the flavour
matrix:
\begin{equation}
{\bf e}^{\pm}(\theta) 
=  \left( 
        \begin{array}{c}
            {\rm e}_s^{\pm}(\theta) \\
            1
        \end{array}
    \right),
\end{equation}
where the $s$ component is,
\begin{equation}
{\rm e}_s^{\pm}(\theta)
= \frac{4 \pm \sqrt{17\cos^4(\theta)
                    +30\cos^2(\theta)\sin^2(\theta)
                    +17\sin^4(\theta)}}
        {\cos^2(\theta)-\sin^2(\theta)}.
\end{equation}
For the dominant mode, Fig. \ref{eigenvector} shows the variation of the
$s$ component as a function of angle.  (For the subdominant mode the same
graph is produced by plotting 1/${\rm e}_s^+(\theta)$.) For momenta lying
along, or near, the diagonals, the dominant mode has an entirely $d$-wave
character.  As we move away from the diagonals and towards the $x$ and $y$
axes, the $s$-wave fraction grows, but even at its maximum value it is an
order of magnitude smaller than the $d$-wave.  We conclude that the
dominant mode is essentially $d$-wave, whilst the subdominant one is
essentially $s$-wave.

Thus, by including both the $s$- and $d$-wave order parameters in the same
theory, as well as the gradient couplings between them, we have been able
to produce a pseudogap state with a non-trivial angular dependence,
qualitatively similar to that of the $d$-wave superconducting state.
Although the angle dependent pseudogap state requires an $s-d$ admixture,
the superconducting state that it eventually gives way to has, for most
intents and purposes, pure $d$-wave symmetry.


\section{Conclusion}


In this paper we  have considered an extended Hubbard model with nearest
neighbour attractive interaction as a phenomenological model for the high
T$_c$ superconducting cuprates.  By introducing a tensorial version of the
Hubbard-Stratonovi{\v c} transformation we were able to derive an effective
Landau-Ginzburg-Wilson action describing both extended $s$- and $d$-wave
superconductivity.  

The saddle point (mean field) approximation to the full action gave the
phase diagram of the model as a function of chemical potential (filling)
and temperature.  Near the centre of the band $d$-wave solutions were found
to be dominant whilst the extended $s$-wave solutions were dominant near
the band edges.  The two flavours of superconductivity were separated from
each other by two second order phase transitions, in between which there
was found to exist a mixed time-reversal symmetry breaking $s \pm id$
phase.  The alternative $s \pm d$ phase was ruled out except in the
presence of orthorhombic distortions.  At large values of the coupling,
non-zero critical temperatures were observed beyond the band edges,
signaling the creation of bosonic bound states.  

Using the Gaussian approximation to the LGW action, in the low density
limit, where quartic terms representing interactions between bosons ought
to be unimportant, we first reproduced the known large $U$ behaviour of the
on-site attractive Hubbard model, in which the bosons acquire a mass
proportional to the coupling, $U$, due to the process of virtual
ionisation.  In the equivalent, large $V$, limit the extended model showed
qualitatively similar behaviour provided only nearest neighbour hopping was
considered.  Introducing a next nearest neighbour hopping, $t^{\prime}$,
allowed diagonal hopping processes, circumventing the need for virtual
ionisation.  It was shown that, in the limit $|t^{\prime}V| \gg t^2$, there
is only one flavour of stable boson, $s$-wave if $t^{\prime}>0$ or $d$-wave
if $t^{\prime}<0$, and that its mass is independent of $V$, just like in a
continuum model.  It was then argued that, for $|t^{\prime}V| \ll t^2$, the
existence of two different bosonic modes, with anisotopic gradient
couplings, could lead to an angle dependent distribution of bosons.  This
was then given a possible interpretation as the cause of the pseudogap
state: two-particle states ``eat'' the single-particle spectral weight in
certain areas of momentum space.  Finally it was shown that, despite the
mixing of the two different boson flavours, the superconducting ground
state would be of predominantly $d$-wave character.


\section*{Acknowledgements}


We would like to thank B.L. Gy{\" o}rffy, J.J. Hogan-O'Neill and
A.M. Martin for useful discussions.  This work was partly supported by
EPSRC under grant number GR/L22454.

\begin{figure}
\centerline{\epsfig{file=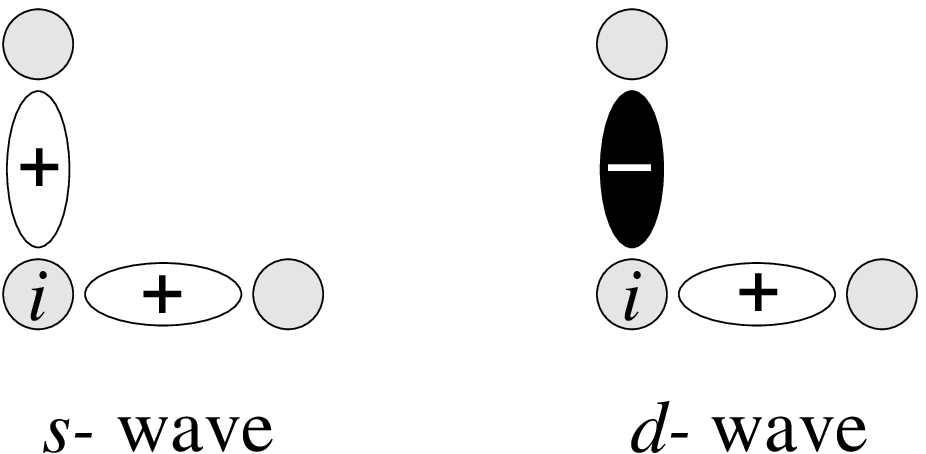,width=10.0cm,angle=0}}
\caption{$s$- and $d$-wave order parameters.}
\label{sd_nodes}
\end{figure}

\end{document}